\documentclass{osa-article}
\usepackage{gensymb}
\usepackage{enumitem}

\journal{oe}

\articletype{Research Article}

\begin{document}

\title{Unified \textit{k}-space theory of optical coherence tomography}
\author{Kevin C. Zhou,\authormark{1,*} Ruobing Qian,\authormark{1} Al-Hafeez Dhalla,\authormark{1} Sina Farsiu,\authormark{1,2} and Joseph A. Izatt\authormark{1,2}}

\address{\authormark{1}Department of Biomedical Engineering, Duke University, Durham, NC 27708\\
\authormark{2}Department of Ophthalmology, Duke University Medical Center, Durham, NC 27708\\
}

\email{\authormark{*}kevin.zhou@duke.edu} 



\begin{abstract*}
We present a general theory of optical coherence tomography (OCT), which synthesizes the fundamental concepts and implementations of OCT under a common 3D $k$-space framework. At the heart of this analysis is the Fourier diffraction theorem, which relates the coherent interaction between a sample and plane wave to the Ewald sphere in the 3D $k$-space representation of the sample. While only the axial dimension of OCT is typically analyzed in $k$-space, we show that embracing a fully 3D $k$-space formalism allows explanation of nearly every fundamental physical phenomenon or property of OCT, including contrast mechanism, resolution, dispersion, aberration, limited depth of focus, and speckle. The theory also unifies diffraction tomography, confocal microscopy, point-scanning OCT, line-field OCT, full-field OCT, Bessel-beam OCT, transillumination OCT, interferometric synthetic aperture microscopy (ISAM), and optical coherence refraction tomography (OCRT), among others. Our unified theory not only enables clear understanding of existing techniques, but also suggests new research directions to continue advancing the field of OCT.
\end{abstract*}

\section{Introduction}
Since its invention nearly three decades ago \cite{huang1991optical}, optical coherence tomography (OCT) has proliferated into a broad class of techniques with a variety of biomedical and clinical applications, such as in ophthalmology, cardiology, dermatology, oncology, and gastroenterology. Even beyond the well known categories of time-domain OCT (TD-OCT) and Fourier-domain OCT (FD-OCT), the field of OCT has evolved to encompass implementations that use a variety of illumination and detection strategies, unified by interferometry with a broadband or low-coherence source. The earliest implementations of OCT were point-scanning OCT systems, which involved scanning a focused spot across the sample, probing one lateral spatial position at a time. Even today, point-scanning OCT remains the most popular form of OCT and is very successful as a clinical standard for ophthalmic imaging and an emerging standard for intravascular and gastroenterological imaging. Shortly thereafter, full-field OCT (FF-OCT) \cite{beaurepaire1998full, dubois2002high, povavzay2006full, subhash2012full, xiao2018vivo, leitgeb2019face} and line-field OCT (LF-OCT) \cite{zuluaga1999spatially, zeylikovich1998nonmechanical, sarunic2006full, nakamura2007high, fechtig2015line} emerged as alternate strategies, which use unfocused or cylindrically (1D) focused light and parallel spatial detection (i.e., a 2D camera or a 1D line camera). Furthermore, apart from Gaussian beams with common mode or confocal detection, other illumination patterns and detection strategies have also been used in OCT, such as Bessel beams with double-pass illumination and detection \cite{ding2002high, lee2008bessel} or with decoupled Gaussian mode detection \cite{leitgeb2006extended, blatter2011extended}, annular pupils for illumination and detection \cite{liu2011imaging, mo2013focus, yu2014depth}, and many other strategies \cite{liu2007binary,lorenser2012ultrathin}. All of these alternative illumination/detection strategies have been used to maintain a high lateral resolution over an extended depth of focus in OCT, compared to the standard Gaussian beam. Other methods have also been proposed to address this issue, notably interferometric synthetic aperture microscopy (ISAM) \cite{ralston2006inverse, ralston2007interferometric}, which computationally corrects the defocus by solving the coherent inverse scattering problem, with different solutions depending on the illumination/detection strategy (e.g., ISAM for Bessel-beam illumination/Gaussian mode detection \cite{coquoz2017interferometric} and FF-OCT \cite{marks2007inverse}). Even more recently, we developed an incoherent angular compounding technique called optical coherence refraction tomography (OCRT) to address this trade-off between depth of focus and lateral resolution by reconstructing an image with isotropic resolution \cite{zhou2019optical, zhou2020spectroscopic}. These various implementations and extensions of OCT would greatly benefit from a unified theoretical treatment that concisely identifies their differences and similarities, their relative advantages and disadvantages, and their relationship to the broader category of coherent imaging. Such a unified theory would also suggest new ways to continue the technological advancement of the field of OCT.

To this end, here, extending our earlier preliminary work \cite{zhou2020resolution}, we present a comprehensive, fully 3D $k$-space analysis of OCT that provides a unified theoretical framework. This theory not only encompasses all of the implementations of OCT mentioned in the previous paragraph, but also explains many fundamental concepts of OCT, including the contrast mechanism, origin of speckle, dispersion, and trade-off between the lateral resolution and depth of focus. Here, $k$-space refers to 3D Fourier space, where $k$ is the customary symbol for representing spatial wavevectors, or $k$-vectors, that compactly denote both the propagation direction of plane waves and the wavelength or wavenumber via the $k$-vector's length. Using principles from the field of Fourier optics, these $k$-vectors serve as a basis for decomposing more complicated waveforms, such as the different types of illumination strategies often employed in OCT (including focused Gaussian beams). As plane waves are the fundamental building blocks for analyzing more complicated systems, we utilize a principle that predicts how a plane wave interacts with a 3D object: the Fourier diffraction theorem, which was developed in the field of diffraction tomography \cite{wolf1969three, lauer2002new, sung2009optical, muller2015theory, horstmeyer2016diffraction, chowdhury2019high, zhou2020diffraction} to reconstruct a sample's 3D refractive index (RI) distribution from a set of diffraction patterns resulting from plane wave illumination from multiple angles. Our $k$-space theoretical treatment builds upon prior excellent reviews and theoretical treatments of OCT \cite{schmitt1999optical, Izatt2015, drexler2014optical,de2017twenty}, including works that analyze OCT in 3D $k$-space \cite{fercher1996optical,  fercher2003optical, coupland2008holography, villiger2010image, ruiz2011depth,  zhou2014inverse, Fercher2015, Adie2015, sentenac2018unified}. However, we advance a unified and comprehensive $k$-space theory of OCT, encompassing a broader range of implementations of OCT and other coherent imaging techniques, as well as being the first one to incorporate speckle as a direct consequence of the band-pass nature of its transfer function.

We note that this paper does not compare TD-OCT and FD-OCT or spectrometer-based FD-OCT and swept-source FD-OCT, as these topics have been treated extensively \cite{leitgeb2003performance, choma2003sensitivity, de2003improved, Izatt2015, de2017twenty}. Rather, from the point of view of $k$-space theory, the differences between these approaches are practical implementation details, insofar as they are different methods of measuring the same optical fields and therefore the same information about the sample. Note that the same cannot be said about point-scanning OCT, FF-OCT, and LF-OCT, which measure slightly different information about the sample, as will become clear.

\section{Towards a 3D $k$-space framework as a general theory of OCT}

Previous theoretical treatments and reviews of OCT (e.g., \cite{schmitt1999optical, Izatt2015, de2017twenty}) are typically centered around low-coherence interferometry, putting forth a 1D $k$-space model (in particular, a distorted version of $k_z$ of our 3D $k$-space model; see Sec. \ref{refocus}) that explains OCT's axial resolving capabilities very accurately. The lateral resolution is then explained using beam focusing, separately from interferometry. While the separability of these explanations is a very good assumption for weakly-focused beams, here, we extend this picture with a fully wave-based, 3D $k$-space model that unites low coherence interferometry and beam focusing under one framework. This is achieved by examining OCT from a more fundamental perspective of the coherent interaction between a plane wave (or superpositions thereof) and a weakly scattering sample. This coherent interaction is straightforwardly visualized in 3D $k$-space via the Ewald sphere, which describes the information obtainable about a sample for a given wavelength and illumination direction, according to the Fourier diffraction theorem (see Sec. \ref{FDT} below). In this work, we show that this $k$-space framework explains and highlights the interdependence among nearly all properties of OCT in a unified manner, particularly the source of contrast, the full 3D point-spread function (PSF) and transfer function (TF), the trade-off between the lateral resolution and depth of focus, and the origin of speckle. We also apply this common $k$-space framework to analyze and compare the TFs of the major implementations of OCT, specifically point-scanning OCT with a Gaussian beam and Bessel beam, LF-OCT, and FF-OCT, as well as coherent confocal microscopy and conventional holography as degenerate cases of OCT. All of these implementations can be thought of as special cases of diffraction tomography in reflection. Finally, we discuss the implications of this theoretical treatment on the limits of speckle reduction and resolution enhancement in OCT.

\begin{figure}
    \centering
    \includegraphics[width=.85\columnwidth]{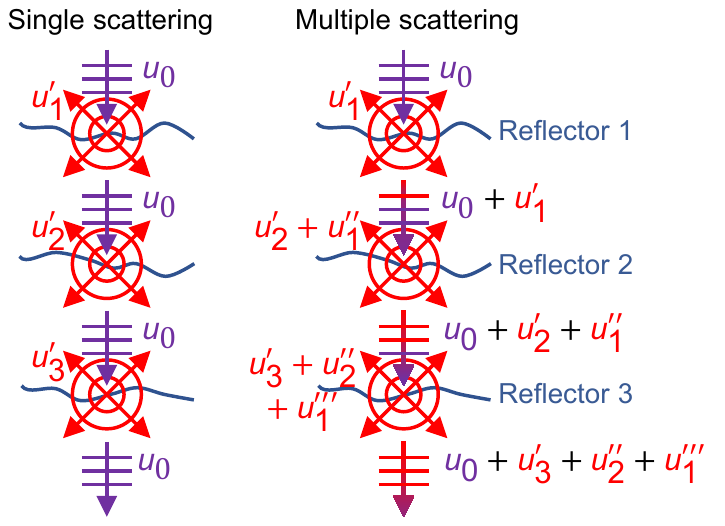}
    \caption{The validity of the first Born approximation in OCT for thick samples. The scattered fields not only contribute to the detected backscattered signal, but may also contribute to forward scattering. $u_0$ is the incident field (purple) and $u_i$ is a scattered field (red), where the subscript indexes which reflector the field was scattered from. The number of primes indicates the number of times that field was scattered. Ideally, if the cumulative RI variation is weak, the incident field is ``regenerated'' upon interaction with each reflector in the sample. If the RI variation is strong, then the incident beam becomes modified by forward scattering from earlier scattering events, resulting in multiple scattering.}
    \label{fig:born}
\end{figure}

\subsection{First Born approximation}
In OCT, one typically makes a weakly or singly scattering assumption about the sample. A common interpretation is in terms of discrete photons and a sample composed of a discrete set of reflectors: an incident photon will interact with exactly one of the reflectors and ignore all the others. Since we are advancing a $k$-space framework, we need to understand this assumption in terms of waves. Thus, we turn to the inhomogeneous, time-independent wave equation (also known as the Helmholtz equation), 

\begin{equation} \label{helmholtz}
    \left(\nabla^2 + k_0^2n_m^2\right) u(\mathbf{r})=-V(\mathbf{r})u(\mathbf{r}),
\end{equation}
where 

\begin{equation}\label{scattering_potential}
V(\mathbf{r})=k_0^2\left(n(\mathbf{r})^2-n_m^2\right) 
\end{equation}
is the sample's scattering potential, which is directly related to its refractive index (RI) distribution $n(\mathbf{r})$, $\mathbf{r}=(x,y,z)$ is the 3D spatial coordinate, $k_0=2\pi/\lambda_0$ is the vacuum wavenumber, and $n_m$ is the background medium RI. Eq. \ref{helmholtz} thus describes the propagation of a wave $u(\mathbf{r})$ through a sample with a spatially varying scattering potential.

As the wave equation only admits closed form solutions to all but the simplest RI distributions (e.g., uniform spheres), the solution in the general case is often obtained through iterative methods or through discrete approximations. One such iterative solution is the Born series, based on a recursive expansion of the Lippmann-Schwinger equation, the integral form of the Helmholtz differential equation (Eq. \ref{helmholtz}). A more detailed explanation of the Born series and the wave equation is beyond the scope of this paper, but has been extensively treated in the literature \cite{wolf1969three, kleinman1990convergent, van1999multiple, sung2009optical, muller2015theory, osnabrugge2016convergent}. 

While the Born series in principle models the general case of multiple scattering, in practice it is unstable and has convergence issues \cite{kleinman1990convergent, osnabrugge2016convergent}. However, truncation of the Born series to only its first term yields a linear equation that permits an interpretable closed form solution given a plane wave illumination (Sec. \ref{FDT}). This is known as the first Born approximation, which states that the emerging field is the superposition of the incident field, $u_{inc}(\mathbf{r})$, and the scattered field $u_{sc}(\mathbf{r})$:
\begin{equation} \label{firstborn}
    u(\mathbf{r})\approx u_{inc}(\mathbf{r})+u_{sc}(\mathbf{r}).
\end{equation}
We can now interpret the meaning of a ``weakly scattering'' or ``singly scattering'' sample in the context of OCT as the condition by which the first Born approximation is valid \cite{chen1998validity}; that is, $u_{sc}$ is much smaller than $u_{inc}$. Note that even though Eq. \ref{firstborn} is expressed in terms of the fields, the validity of the first Born approximation is a property of the sample, not the illumination.

The first Born model is a reasonable assumption in OCT, as the backscattered signals are typically several orders of magnitude smaller than incident beam for most biological samples. Note that this assumption does not necessarily place a limit on the sample's thickness, but rather the cumulative RI variation across the sample depth. For example, while a sample with very small RI variation can be thicker, a sample with high RI variation only satisfies the first Born approximation if it is thin. With enough cumulative RI variation, the sample becomes multiply scattering. Fig. \ref{fig:born} illustrates this point intuitively with a multi-layer Born model \cite{chen2019multi}, a multiple scattering model developed for diffraction tomography that divides the thick sample into layers within each of which the first Born approximation applies (with the caveat that the multi-layer Born model is not equivalent to the Born series, as the former does not consider bidirectional interaction among the layers). From this interpretation, in the multiply scattering case, the incident field on a deep layer within the sample has been aberrated through cumulative interactions with shallower RI variations. In fact, OCT is routinely operated outside of the first Born approximation, as evidenced by shadowing or attenuation at greater depths, which is not predicted by the first Born model. This suggests that OCT does not necessarily fail when the first Born assumption is broken, but rather as long as the field incident at a deep structure is not completely random (i.e., it is primarily forward scattering and therefore contains some memory of the incident field), one can still obtain depth-resolved measurements, at the cost of signal-to-noise ratio (SNR) due to inefficient back-coupling into the fiber, which acts as a spatial mode filter in the case of point-scanning OCT, or due to cross-talk in the case of FF-OCT. 

However, for the ensuing $k$-space theoretical treatment, we will assume validity of the first Born approximation.

\begin{figure}
    \centering
    \includegraphics[width=.85\columnwidth]{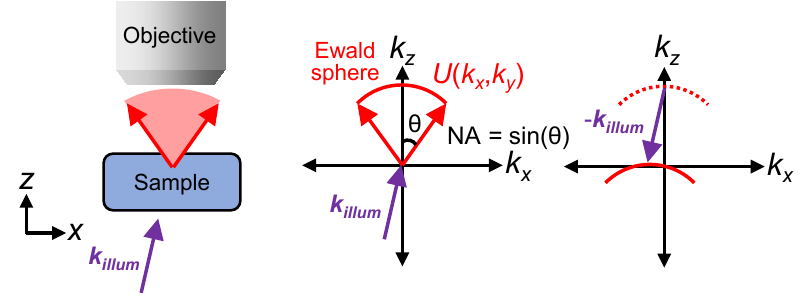}
    \caption{An incident plane wave, denoted by $\mathbf{k_{illum}}$, illuminates a sample that obeys the first Born approximation, which scatters light in potentially every direction (red circle, the Ewald sphere). As in Fig. \ref{fig:born}, purple corresponds to incident fields and red corresponds to scattered fields. Only the field contained within the solid angle covered by the objective lens, $U(x,y)\overset{\mathcal{F}}{\leftrightarrow}U(k_x,k_y)$, is measured. In $k$-space, the 2D measured field corresponds to the surface of the Ewald sphere. To obtain the $k$-space coverage according to the illumination geometry, the origin-centered partial Ewald sphere is translated by subtracting out $\mathbf{k_{illum}}$.}
    \label{fig:FDT}
\end{figure}

\subsection{Contrast mechanism of OCT}
As light propagation throughout a sample is dictated by Eq. \ref{helmholtz}, the source of contrast in OCT is the spatially varying scattering potential, which directly relates to the sample's RI distribution. As such, properties such as scattering coefficients, scattering phase functions, and anisotropy factors commonly used to characterize bulk scattering properties of tissue are simply higher-level descriptors that are based on the more fundamental scattering potential or RI variation. Note that scattering potential and RI can be complex-valued, meaning in theory the spatial variation in absorption can also influence how light propagates through the sample. However, as we will show, not all types of RI variation will produce a detectable signal in OCT, as only those which produce backscattering can contribute to OCT contrast. To appreciate what properties of the sample's RI distribution that OCT is sensitive to, and conversely how the RI distribution affects an input illumination field, we turn to the FDT.

\subsection{Fourier diffraction theorem} \label{FDT}
The FDT is a fundamental theorem for diffraction tomography that relates a sample's 3D scattering potential to the complex 2D diffraction pattern of a plane wave, viewed from an arbitrary direction \cite{muller2015theory,wolf1969three}. Its relevance to OCT has also been pointed out \cite{fercher2003optical,Fercher2015}. The FDT can be thought of as the wave analog of the projection-slice theorem (also known as the Fourier slice theorem), which assumes a ray model and is commonly used for X-ray computed tomography (CT). In the geometric optics limit ($\lambda\rightarrow0$), the two theorems converge. 

Specifically, consider a sample with scattering potential $V(x,y,z)$ and a monochromatic plane wave governed by the wavevector $\mathbf{k_{illum}}=(k_{illum, x},k_{illum, y},k_{illum, z})$, which describes the direction of the field as well as its wavelength or wavenumber, i.e., $|\mathbf{k_{illum}}|=k_0$. Without loss of generality, assume that the sample is at the origin in a 3D Cartesian space and that we are interested in the 2D diffraction pattern, $U(x,y)$, at $z=0$ in the $xy$ plane (or a conjugate plane thereof). The FDT states 
\begin{equation} \label{FDT_eq}
    \widetilde{U}(k_x,k_y)\propto\widetilde{V}\left(\left(k_x,k_y,\sqrt{k_0^2-k_x^2-k_y^2}\right)-\mathbf{k_{illum}}\right),
\end{equation}
where the tildes denote the Fourier transforms, $\widetilde{V}\left(k_x,k_y,k_z\right)=\mathcal{F}_{3D}\{V(x,y,z)\}$ and $\widetilde{U}\left(k_x,k_y\right)=\mathcal{F}_{2D}\{U(x,y)\}$. The argument of $\widetilde{V}$ describes a displaced Ewald spherical shell in $k$-space with radius $k_0$ and center $\mathbf{k_{illum}}$ (Fig. \ref{fig:FDT}). In practice, only a small solid angle surrounding the $k_z$-axis is accessible due to the limited numerical aperture (NA) of the objective lens. This partial spherical shell in $k$-space can be thought of as the 3D transfer function (TF) of the sample's scattering potential for monochromatic plane wave illumination and full-field collection. While this TF is relatively modest in its $k$-space coverage, it forms the basic building block of general coherent imaging modalities that may use angular diversity (the angular spectrum) and wavelength diversity (the source spectrum) to construct larger TFs.

Just as the FDT is a fundamental theorem for diffraction tomography, it is also a fundamental theorem for OCT, as a reflection-mode coherent imaging modality. In the following sections, we apply the FDT to a variety of coherent imaging modalities and derive their respective transfer functions of the information in the sample's scattering potential.

\begin{figure}
    \centering
    \includegraphics[width=.85\columnwidth]{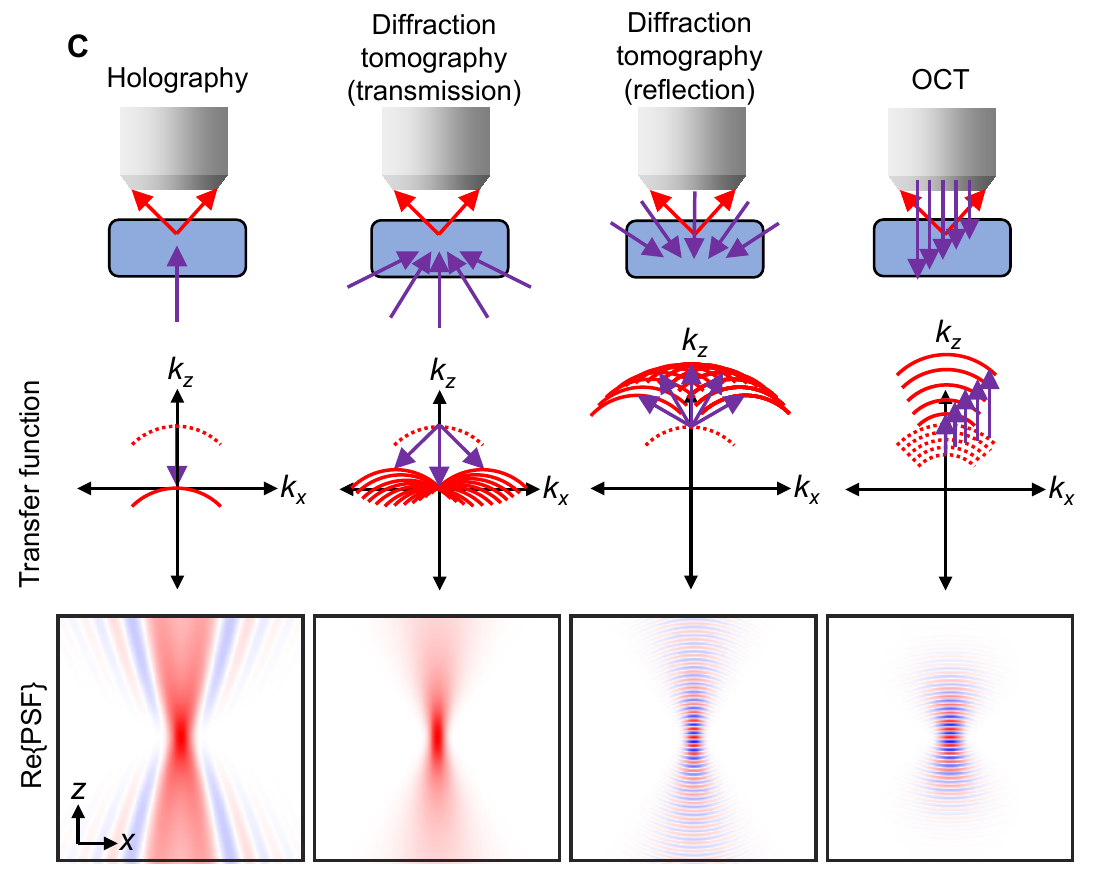}
    \caption{FDT can be used to derive the transfer functions (TFs) and PSFs of (from left to right): holography (single illumination angle), diffraction tomography (DT) in transmission (multi-angle illumination with fixed collection in transmission), DT in reflection (multi-angle illumination with fixed collection in reflection), and FF-OCT (single-angle illumination in reflection with multiple wavelengths).}
    \label{fig:TF_PSF_overview}
\end{figure}

\section{Transfer functions of various coherent imaging modalities} \label{transfer_functions}
\subsection{Holographic microscopy} \label{holography}
Perhaps the simplest case is holographic microscopy (or simply, holography \cite{gabor1948new, schnars2015digital}), which is a useful starting point to understand the FDT and is the building block for our extension of the 3D $k$-space formalism to OCT implementations. Holography uses monochromatic plane wave illumination, potentially from any angle, to interrogate the sample and images the emerging diffracted field onto a 2D film or camera with phase-sensitive detection (e.g., by use of an off-axis reference or an on-axis reference with multiple phase shifts). Depending on the direction of illumination relative to collection direction (assumed to be the $z$-axis), the partial Ewald sphere will be shifted to a different position in $k$-space. For reflective geometries, holography probes high spatial frequencies, while for transmissive geometries, holography probes low frequencies. Holography always has poor axial sectioning capabilities in 3D samples, as the TF is infinitely thin in the axial direction. For thin, 2D samples to which holography is often applied, the lack of axial width in the TF is not a concern, as a thin sample's scattering potential spectrum in $k$-space is invariant in the $k_z$ direction (i.e., the Fourier transform of a delta function is a constant). This 2D limiting case is known variously as quantitative phase imaging (QPI \cite{park2018quantitative}), in which the 3D structure of the Ewald sphere can be ignored and regarded as circles in the $k_xk_y$ plane.

Although holography has poor axial sectioning (Fig. \ref{fig:TF_PSF_overview}), because the full 2D field of the scattered wave is measured, we can use the Fresnel diffraction kernel to digitally propagate the field to, in principle, any axial position within the 3D sample. While out-of-plane features still appear, this property of holography has an interesting implication -- in theory, it doesn't matter where the camera is placed after the sample, whether at an image plane, Fourier plane, or directly next to the sample (in practice, putting the camera at an image plane is better to improve SNR). As we will see, this property implies that, at first glance, OCT should not have a limited depth of focus, which we discuss in Sec. \ref{refocus} below.

\subsection{Diffraction tomography (multi-angle holography)}
Diffraction tomography takes holography a step further and uses angular diversity to synthesize a wider TF \cite{wolf1969three, lauer2002new, sung2009optical, muller2015theory, horstmeyer2016diffraction, chowdhury2019high, zhou2020diffraction}. One strategy is to use a fixed sample and detector, but to vary the illumination angle. Another is to have a fixed detector and illumination geometry, but to rotate the sample. These strategies can be implemented in transmission or reflection mode, with different synthesized TFs, which are depicted in Fig. \ref{fig:TF_PSF_overview}. Clearly, with angular diversity, the achievable TFs are far more substantial than with holography. We also note that reflective geometries tend to generate band-pass TFs, while transmissive geometries tend to generate low-pass TFs. This means that transmissive geometries are better suited for measuring quantitative RI values, while reflective geometries are generally only sensitive to changes in RI. Finally, just as with holography, this $k$-space analysis makes no reference to the depth of focus of the imaging lens, and thus the reconstruction volume is not theoretically limited when using high-NA objectives, but rather limited by SNR and the validity of the first Born approximation.

\begin{figure}
    \centering
    \includegraphics[width=.85\columnwidth]{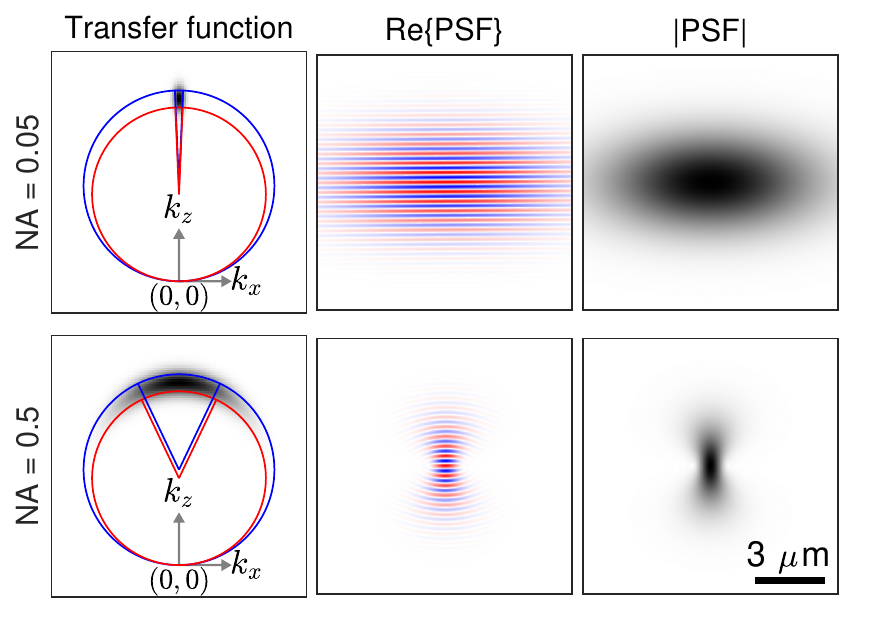}
    \caption{Simulated TFs and PSFs for FF-OCT at $\lambda_0=820$ nm at a low NA (top row) and high NA (bottom row). The red and blue circles represent the Ewald spheres of the wavenumbers corresponding to the FWHM of the source spectrum. Similarly, the red and blue wedges correspond to the FWHM angular range of the focused Gaussian beam. In the low-NA case (top row), both the TF and PSF are approximately separable into their Gaussian axial and lateral components.}
    \label{fig:FFOCT_TF_PSF}
\end{figure}

\subsection{Full-field OCT} \label{FFOCT}
FF-OCT uses a fixed, reflective imaging geometry from the same aperture, but uses a broadband source instead of monochromatic illumination \cite{beaurepaire1998full, dubois2002high, povavzay2006full, subhash2012full, leitgeb2019face}. As with the previous examples, the interference pattern is recorded with a 2D camera. Here, we will assume a wavelength-swept source system to simplify the explanation, but the same analysis holds for a time-domain FF-OCT system, which can be conceptually decomposed to monochromatic plane waves (cf., Fourier optics). Thus, FF-OCT can be thought of as performing reflective holography at multiple wavelengths. The Ewald sphere has a different radius for different wavelengths (i.e., $|\mathbf{k_{illum}}|=2\pi/\lambda_{illum}$), so a continuous sweep synthesizes a continuous 3D band-pass (Fig. \ref{fig:TF_PSF_overview}). More precisely, given a Gaussian spectrum centered at $k_0$ with a standard deviation width parameter of $\sigma_k$, and an imaging objective with $\mathrm{\mathit{NA}}=\sin(\sqrt{2\ln{2}}\sigma_{\theta})$, defined such that the half width at half maximum (HWHM) collection angle is $\sqrt{2\ln{2}}\sigma_{\theta}$, the TF of OCT is
\begin{equation}
    H_{\mathrm{\mathit{FFOCT}}}(k_x, k_y, k_z)\propto\exp\left(-\frac{(k_r-2k_0\cos(k_\theta))^2}{8\sigma_k^2\cos^2(k_\theta)}\right)\exp\left(-\frac{2k_\theta^2}{\sigma_\theta^2}\right),
    \label{OCT_TF}
\end{equation}
where $k_r=\sqrt{k_x^2+k_y^2+k_z^2}$ and $k_\theta=\cos^{-1}(k_z/k_r)$ are the 3D $k$-space coordinates in spherical coordinates (the azimuthal angle is not needed because the TF is symmetric about the $k_z$-axis). Although the two exponential factors in Eq. \ref{OCT_TF} are coupled, the first factor is more directly related to the axial resolution, while the second factor is more related to the lateral resolution. Thus, the FF-OCT TF is a solid angle with a half angle of $\sigma_\theta/2$ (i.e., half that of the objective) and centered at $k_z=2k_0$, where $k$-space theory recovers the factor of 2 attributed to the round trip of the input beam using conventional OCT theory. One caveat with Eq. \ref{OCT_TF} is that in reflection, only the upper hemisphere of the Ewald sphere is detected, as the lower hemisphere corresponds to transmission in the opposite direction (i.e., the $-k_z$-axis). Example TFs are plotted in Fig. \ref{fig:FFOCT_TF_PSF}.

In the low-NA limit ($\sigma_\theta\rightarrow0$),
Eq. \ref{OCT_TF} can be approximated by

\begin{equation}
    H_{\mathrm{\mathit{FFOCT}}}(k_x, k_y, k_z)\approx \exp\left(-\frac{(k_z-2k_0)^2}{8\sigma_k^2}\right)\exp\left(-\frac{2k_{xy}^2}{\sigma_{k_{xy}}^2}\right),
    \label{OCT_TF_approx}
\end{equation}
where $k_{xy}=\sqrt{k_x^2+k_y^2}\approx2k_0k_\theta$ and $\sigma_{k_{xy}}=2k_0\sigma_\theta$ (Fig. \ref{fig:FFOCT_TF_PSF}, first row).
In this limit, the TF is elliptical and therefore separable into its axial and lateral components, both of which are Gaussian. Thus, we can compute the axial and lateral PSFs analytically, given the Fourier transform pair, $\exp(-x^2/(2\sigma^2)) \overset{\mathcal{F}}{\leftrightarrow}\exp(-\sigma^2k^2/2)$.
The axial PSF is thus
\begin{equation}\label{axial_PSF}
    \mathrm{\mathit{psf}}_z(z)\propto \exp\left(-\frac{z^2}{2\sigma_z^2}\right)\exp(j2k_0z),
\end{equation}
where the axial Gaussian width parameter and the axial resolution are given by
\begin{equation}\label{axial_resolution}
    \sigma_z=\frac{1}{2\sigma_k} \implies
    \delta z_{\mathrm{\mathit{FWHM}}}=2\sqrt{2\ln(2)}\sigma_z=\frac{\sqrt{2\ln(2)}}{\sigma_k}=\frac{2\ln(2)}{\pi}\frac{\lambda_0^2}{\Delta \lambda}\approx
    0.44\frac{\lambda_0^2}{\Delta \lambda},
\end{equation}
where $\Delta \lambda$ is the FWHM bandwidth of the source in wavelength and $\lambda_0$ is the center wavelength. This $k$-space derivation of the axial resolution is consistent with conventional OCT theory \cite{Izatt2015}. Note that Eq. \ref{axial_PSF} is identical to the complex coherence function, except for a factor of 2 in the argument of the complex exponential. Similarly, if we analyze the lateral component of Eq. \ref{OCT_TF_approx}, we obtain the lateral PSF,
\begin{equation} \label{lateral_psf}
    \mathrm{\mathit{psf}}_{xy}(x,y)\propto \exp\left(-\frac{x^2+y^2}{2\sigma_{xy}^2}\right),
\end{equation}
with lateral Gaussian width parameter and lateral resolution
\begin{equation} \label{lateral_resolution}
    \sigma_{xy}=\frac{2}{\sigma_{k_{xy}}}=\frac{1}{k_0\sigma_\theta}\implies
    \delta xy_{\mathrm{\mathit{FWHM}}}=2\sqrt{2\ln(2)}\sigma_{xy}=
    \frac{2\ln(2)}{\pi}
    \frac{\lambda_0}{\mathrm{\mathit{NA}}}
    \approx
    0.44\frac{\lambda_0}{\mathrm{\mathit{NA}}},
\end{equation}
also consistent with conventional OCT theory. We emphasize that this separability between and the availability of analytical expressions for the axial and lateral components are only possible when the approximation in Eq. \ref{OCT_TF_approx} is valid (i.e., low NAs). In general, however, the FF-OCT TF is governed by Eq. \ref{OCT_TF}, in which there is a coupling between the axial and lateral dimensions. In other words, the mechanisms for axial and lateral resolution are \textit{not} independent.

Interestingly, note the parallels in the prefactors in the Eqs. \ref{axial_resolution} and \ref{lateral_resolution} when expressed in terms of wavelength. Thus, in addition to the usual interpretation of the axial resolution being governed by the source bandwidth, we can interpret the the lateral resolution as governed by an angular bandwidth. This further suggests that the general 3D $k$-space framework treats axial and lateral resolution on equal footing.

The consequence of the band-pass nature of this TF is that OCT is only sensitive to rapid changes in the scattering potential or RI primarily in the axial direction. Quantitative RI values are close to the $k$-space origin and thus under normal circumstances, OCT cannot measure absolute RI. OCT also cannot detect gradual RI gradients (e.g., gradient index lenses) as these manifest as low-frequency components outside of the band-pass. Likewise, while horizontal edges (i.e., parallel to the $xy$ plane) are visible in OCT, vertical edges are not (see Sec. \ref{coherence_gating} for a more detailed explanation). More generally, the sensitivity of OCT to tilted edges depends on the magnitude Fourier components within the OCT band-pass.

\begin{figure}
    \centering
    \includegraphics[width=.9\columnwidth]{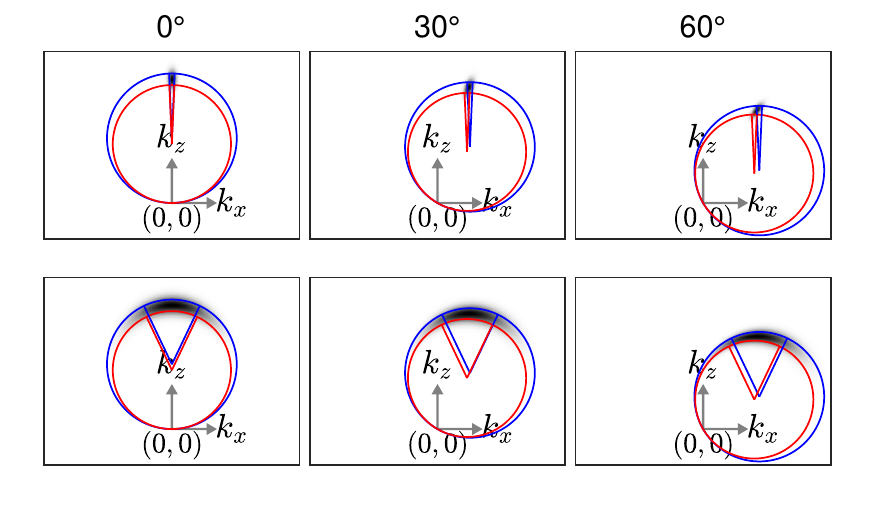}
    \caption{Example FF-OCT TFs with illumination directions at 0\degree, 30\degree, and 60\degree\ with respect to the $-k_z$ direction, for $\lambda_0=820$ nm, NA=0.5, and $\Delta\lambda=300$ nm. Note that the TF not only shifts position with illumination angle, but also changes shape.}
    \label{fig:tilted_illum}
\end{figure}

\subsection{Full-field OCT with off-axis illumination}
As a stepping stone to understanding how confocal microsocpy and point-scanning OCT fit in this $k$-space framework, we first analyze FF-OCT using different illumination angles relative to the objective position. In other words, we generalize Eq. \ref{OCT_TF}, which assumes $\mathbf{k_{illum}}=(0,0,-k_0)$, to allow $\mathbf{k_{illum}}=(k_{illum, x},k_{illum, y},k_{illum, z})$ to specify an arbitrary plane wave. For convenience, we define the following quantities:
\begin{itemize}
    \item $\mathbf{k_{illum}}=(k_{illum,r}=k_0,k_{illum,\theta},k_{illum,\phi})$, the spherical coordinate representation (where $\theta$ and $\phi$ correspond to inclination angle and azimuthal angles, respectively),
    \item $\mathbf{k_{illum,\theta/2}}=(k_0,k_{illum,\theta/2},k_{illum,\phi})$,
    \item $k'_{\theta}=\cos^{-1}\left(\frac{\mathbf{k}\cdot\mathbf{k_{illum}}}{k_rk_0}\right)$, the angle between $\mathbf{k}$ (the 3D $k$-space coordinates) and $\mathbf{k_{illum}}$, and
    \item $k'_{\theta,1/2}=\cos^{-1}\left(\frac{\mathbf{k}\cdot\mathbf{k_{illum,1/2}}}{k_rk_0}\right)$, the angle between $\mathbf{k}$ and $\mathbf{k_{illum,1/2}}$.
\end{itemize}
Then, the FF-OCT TF for an arbitrary illumination wavevector $\mathbf{k_{illum}}$ is given by

\begin{equation}
    H_{\mathrm{\mathit{FFOCT}}}\left(k_x, k_y, k_z;\mathbf{k_{illum}}\right)\propto\exp\left(-\frac{(k_r-2k_0\cos(k'_\theta))^2}{8\sigma_k^2\cos^2(k'_\theta)}\right)\exp\left(-\frac{2k'^2_{\theta,1/2}}{\sigma_\theta^2}\right),
    \label{OCT_TF_general}
\end{equation}
noting that this equation reduces to Eq. \ref{OCT_TF} for $\mathbf{k_{illum}}=(0,0,-k_0)$. Fig. \ref{fig:tilted_illum} shows example TFs for several illumination angles and different bandwidth/NA combinations. In particular, rotating the illumination causes the TF to shift laterally in $k$-space, similarly to the case monochromatic diffraction tomography Fig. \ref{fig:TF_PSF_overview}, except with a change in the shape of the FF-OCT TF due to the difference in curvature of the Ewald spheres for different illumination colors.

Another reason why this imaging geometry might be useful to consider is that it is another approach to coherently enhance the lateral resolution of FF-OCT, which to our knowledge has not yet been demonstrated experimentally. We compare this approach to ISAM in Sec. \ref{coherent_approaches}.

\begin{figure}
    \centering
    \includegraphics[width=.85\columnwidth]{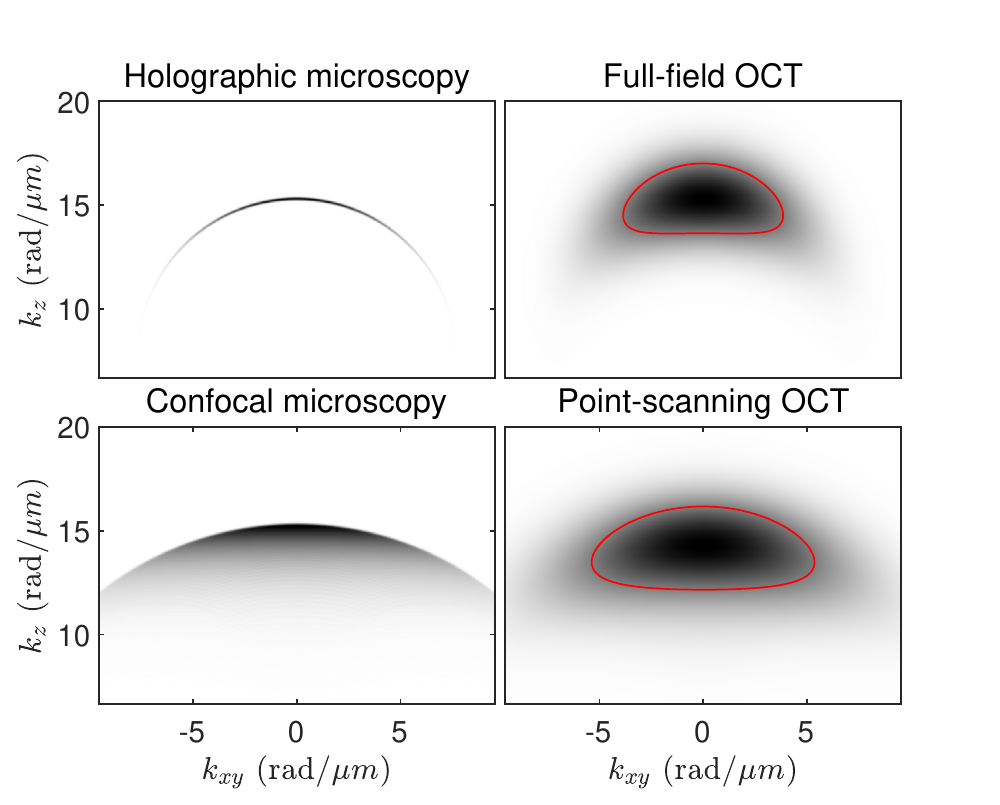}
    \caption{Comparison of TFs for monochromatic holographic microscopy (Sec. \ref{holography}), FF-OCT (Sec. \ref{FFOCT}), monochromatic reflective confocal microscopy (Sec. \ref{confocal}), and point-scanning OCT (Sec. \ref{point_scanning_OCT}), for $\lambda_0=820$ nm, NA=0.5, and, for the right column, $\Delta\lambda=300$ nm. The red curves are half-max contours. Note that confocal microscopy has axial sectioning, evident from the axial extent of its TF. Both confocal microscopy and point-scanning OCT obtain lateral resolution enhancement over their wide-field analogs.}
    \label{fig:four_TFs_high_NA}
\end{figure}

\begin{figure}
    \centering
    \includegraphics[width=.85\columnwidth]{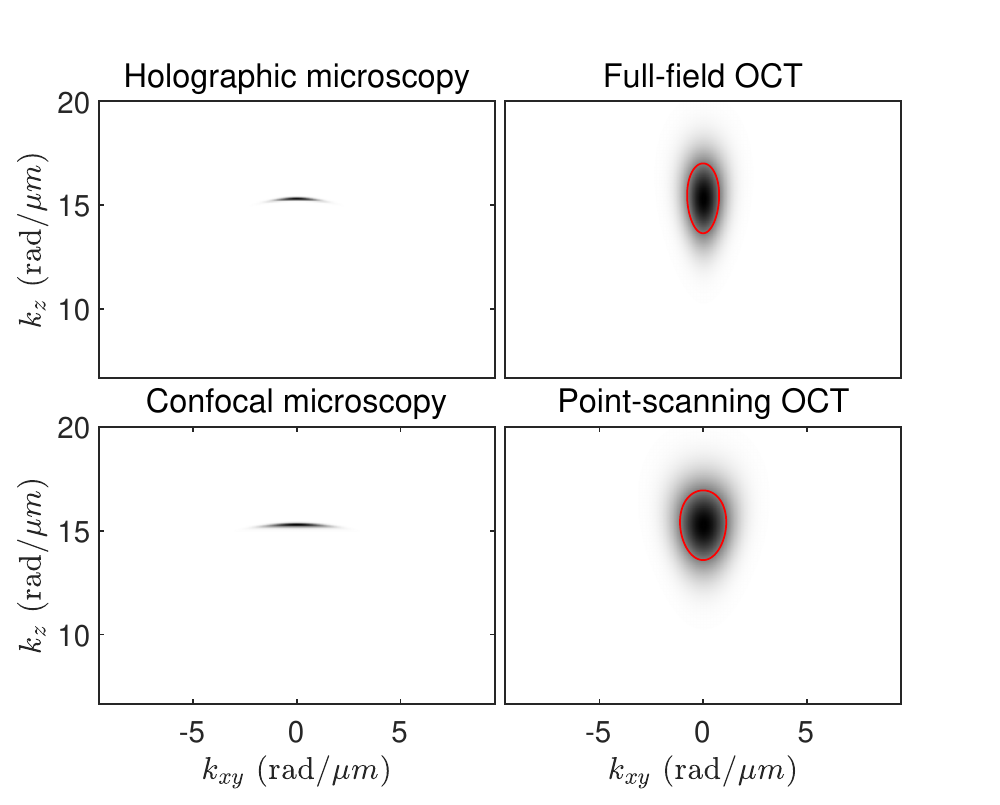}
    \caption{The same comparison of TFs as in Fig. \ref{fig:four_TFs_high_NA}, except at a lower NA of 0.1 (the other parameters are the same, with $\lambda_0=820$ nm and $\Delta\lambda=300$ nm for the right column). In this regime, the confocal microscopy TF has much less axial sectioning compared to the high-NA regime. Further, the FF-OCT and point-scanning OCT TFs are more separable between their axial and lateral components (i.e., better approximated by the product of orthogonal Gaussians).}
    \label{fig:four_TFs_low_NA}
\end{figure}

\subsection{Point-scanning OCT} \label{point_scanning_OCT}
For the coherent imaging methods considered up until now, we have assumed an array detector (e.g., a 2D camera) with plane wave illumination. However, the most common implementation of OCT is with point-scanning, that is, raster-scanning a focused point and using a point detector, thereby encoding space in time. While the FDT (Eq. \ref{FDT_eq}) deals with plane wave illumination, we can synthesize any input illumination wavefront given its angular spectrum (the complex field amplitude as a function of the illumination $k$-vector). For this analysis, we assume a focused Gaussian beam with the same illumination NA as imaging NA, as typically enforced by an input fiber acting as the illumination source and collection aperture. The angular spectrum at the focus is given by
\begin{equation}
    E(k_x, k_y)=E_0\exp\left(-\frac{k_x^2+k_y^2}{k_0^2\mathrm{\mathit{NA}}^2}\right),
\end{equation}
with the caveat that this equation includes evanescent fields, which, however, are negligible given that the NA is typically small in OCT.

Thus, to calculate the TF of Gaussian-beam-illuminated, point-scanning OCT, we perform a superposition of different tilted Ewald spheres (Eq. \ref{OCT_TF_general}), weighted by the complex amplitude of the plane wave given by the illumination angular spectrum:
\begin{equation}\label{TF_OCT_point_scan}
\begin{split}
    &H_{OCT}^{scan}(k_x,k_y,k_z)=\\
    &\iint\displaylimits_{k_{i,x}^2+k_{i,y}^2<k_0^2}
    E(k_{i,x},k_{i,y})H_{\mathrm{\mathit{FFOCT}}}\left(k_x, k_y, k_z;k_{i,x},k_{i,y},\sqrt{k_0^2-k_{i,x}^2-k_{i,y}^2}\right)dk_{i,x}dk_{i,y},
\end{split}
\end{equation}
where integration is over the domain of non-evanescent, propagating waves. Thus, point-scanning OCT (and coherent confocal microscopy) are similar to performing diffraction tomography over the equivalent angular range. The integral in Eq. \ref{TF_OCT_point_scan} is akin to a convolution integral, except that the TF changes shape as the illumination angle is swept. Thus, it does not in general yield an analytical solution, so it needs to be evaluated numerically (Fig. \ref{fig:four_TFs_high_NA}). We can see that point-scanning OCT has \textasciitilde{}$\sqrt{2}$ improvement in lateral resolution over FF-OCT with the same NA due to confocal gating (note, however, that the frequency cutoff doubles, which is clearer when considering a non-Gaussian TF with a hard cutoff), as well as an improvement in axial resolution for larger NAs, as we are entering the optical coherence microscopy (OCM) regime in which the axial confocal gate and coherence gate become comparable to each other \cite{izatt1994optical}. Note also in these higher-NA cases that the radius of curvature of the point-scanning OCT TF appears to have increased compared to the FF-OCT TF, whose implications we discuss in Sec. \ref{refocus}. Fig. \ref{fig:four_TFs_low_NA} shows the same comparison but at a lower NA, at which the OCT TFs are more separable into their axial and lateral components. Here, we again see the improvement in lateral resolution of point-scanning OCT over FF-OCT. However, the axial resolution is similar, due to the weaker confocal gate than in the higher-NA case.

We also note that Eq. \ref{TF_OCT_point_scan} is general and can be used for any input illumination profile, as long as the angular spectrum is known. In particular, we can use Eq.\ref{TF_OCT_point_scan} in later sections to derive the TFs for other spatial scanning techniques, such as confocal microscopy (Sec. \ref{confocal}) and LF-OCT (Sec. \ref{line_field}). We can also simulate the effects of aberration on the TFs by imparting a 2D phase profile in the angular spectrum.

\subsection{Confocal microscopy} \label{confocal}

We can also use Eq. \ref{TF_OCT_point_scan} to compute the TF of coherent confocal microscopy by setting $\sigma_k$ to a very small value (i.e., the monochromatic limit). The results are shown in Figs. \ref{fig:four_TFs_high_NA} and \ref{fig:four_TFs_low_NA}, which are consistent with previous derivations of TFs for confocal microscopy assuming circular apertures \cite{sheppard1990three,sheppard1992significance,sheppard1994three}, which describe the TF as the convolution of two Ewald spheres. Note that although Eq. \ref{TF_OCT_point_scan} is not in general a convolution integral, it is for each individual wavenumber ($\sigma_k\rightarrow0$). In both transmission and reflection, we observe the $\sqrt{2}$ lateral resolution enhancement, but we can also see the optical sectioning effects at higher NAs. Thus, we can conclude that the origin of axial confocal gating is the curvature of the Ewald sphere, which is only appreciable at high NAs, as the axial extent of the TF would not change if there were no curvature. This is why the depth of focus is not significantly reduced in point-scanning OCT, which typically uses low NA beams that do not have significant curvature in $k$-space. Finally, we can see the change in radius of curvature of the TF more clearly in confocal microscopy than in point-scanning OCT, which we discuss in Sec. \ref{refocus}.

\begin{figure}
    \centering
    \includegraphics[width=.85\textwidth]{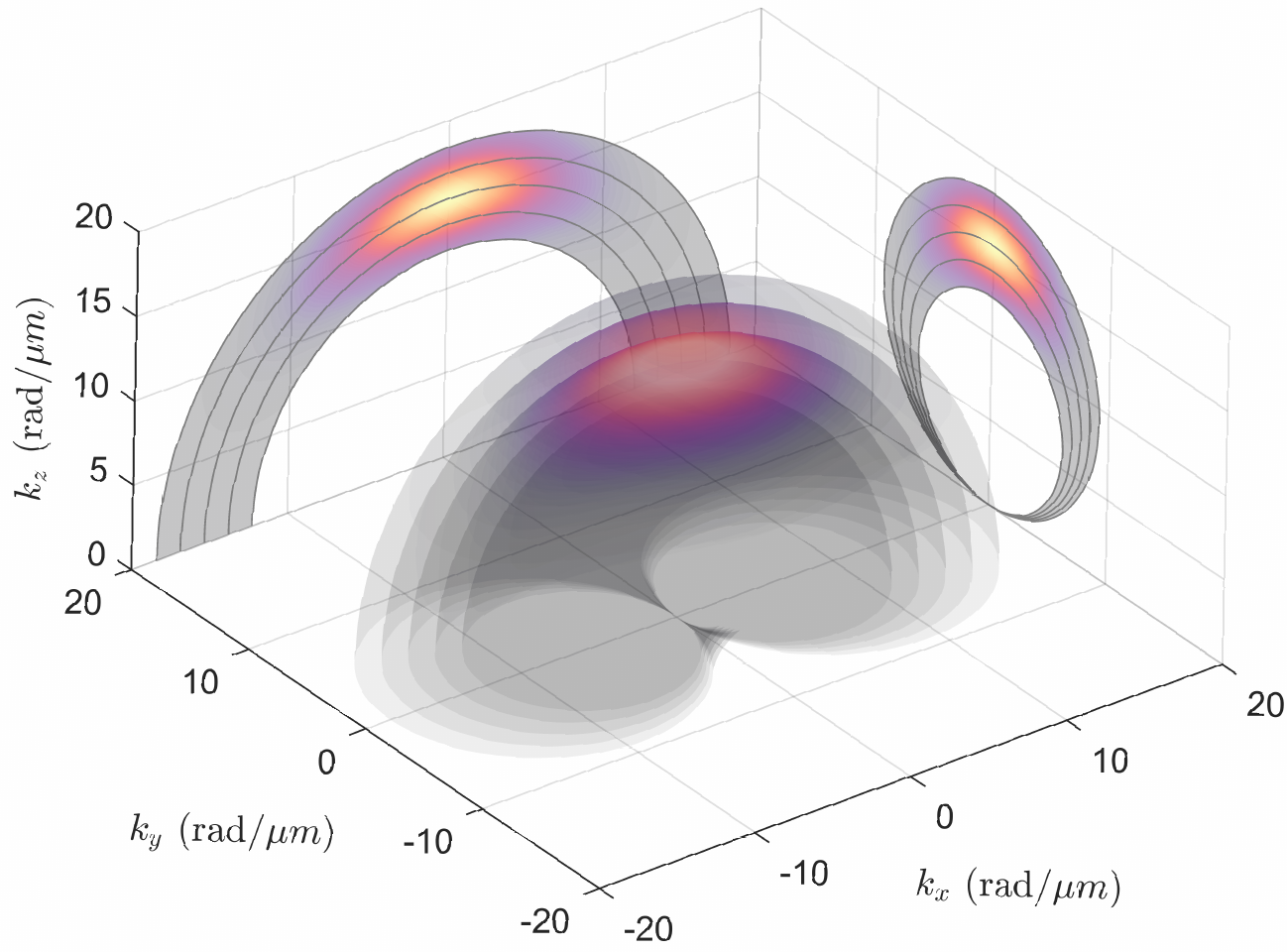}
    \caption{The TF of LF-OCT is asymmetric in 3D $k$-space and follows the curvature of a horn torus. For illustrative purposes, a discrete number of surfaces (5) are shown, corresponding to different wavenumbers. The projections onto the $k_xk_z$- and $k_yk_z$-planes correspond to the TFs of point-scanning and FF-OCT, respectively (for a LF-OCT system focusing in the $x$ dimension). The surfaces depicted are the resampling surfaces to obtain depth-invariant resolution in ISAM (Sec. \ref{refocus}).}
    \label{fig:LFOCT}
\end{figure}

\subsection{Line-field OCT} \label{line_field}
In LF-OCT, the illumination beam is focused only in one dimension \cite{zuluaga1999spatially, zeylikovich1998nonmechanical, sarunic2006full, nakamura2007high, fechtig2015line}. Thus, LF-OCT behaves like a point-scanning OCT system along the focused dimension, while behaving like a FF-OCT system along the unfocused dimension. In other words, we can write the angular spectrum of the LF-OCT illumination at the focus as
\begin{equation} \label{LFOCT_angular_spectrum}
    E(k_x, k_y)=E_0\exp\left(-\frac{k_x^2}{k_0^2\mathrm{\mathit{NA}}^2}\right)\delta(0, k_y),
\end{equation}
where $\delta(x,y)$ is the 2D Dirac delta function, and the illumination beam is focused in the $x$ dimension, but not $y$. It is tempting to regard the $x$ and $y$ dimensions of the TF of LF-OCT as separable into the TFs for FF-OCT and point-scanning OCT, respectively; however, upon inspection of the Eq. \ref{TF_OCT_point_scan}, we note that the integral is not separable because while Eq.\ref{LFOCT_angular_spectrum} is separable, Eq. \ref{OCT_TF_general} is not separable, except at low NAs. In the general case, each wavenumber component of the TF of LF-OCT has the shape of a horn torus (a torus without a hole in the center) (Fig. \ref{fig:LFOCT}); that is, the surface described by revolving about the $k_y$-axis a circle with radius $k_{illum}$ and centered at $k_{illum}$ in the $k_xk_z$-plane (i.e., the cross-section of the Ewald sphere).

\begin{figure}
    \centering
    \includegraphics[width=.85\columnwidth]{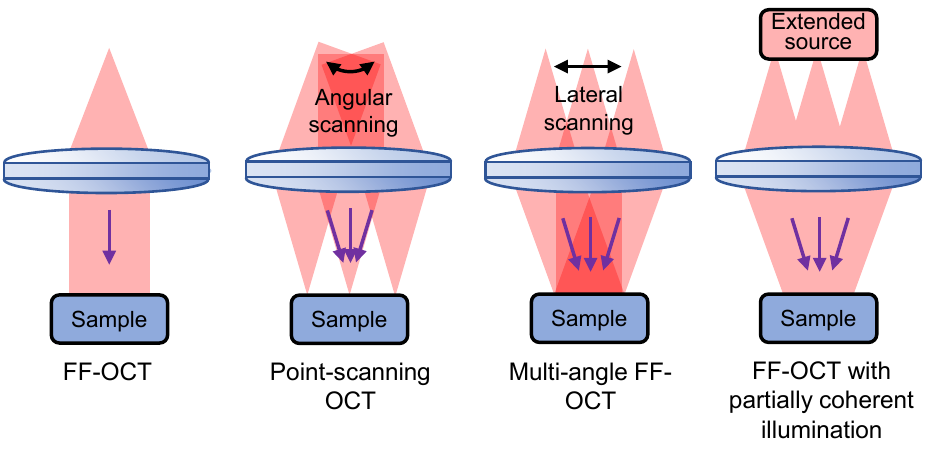}
    \caption{Comparison of three approaches for doubling the lateral frequency cutoff over FF-OCT by increasing illumination $k$-vector diversity, thus attaining the TF in the lower right panel of Fig. \ref{fig:four_TFs_high_NA}: point-scanning OCT of a focused beam, performing FF-OCT at multiple angles, and FF-OCT with partially spatially coherent illumination via an extended source.}
    \label{fig:resolution_doubling}
\end{figure}

\subsection{Full-field OCT with partially spatially coherent illumination}
\label{partial_coherence}
Up until now, we have considered fully spatially coherent plane-wave illumination when deriving the OCT TFs. However, in FF-OCT it is common to use spatially incoherent illumination to mitigate cross-talk issues due to full-field detection \cite{karamata2004spatially, marks2009partially, dhalla2010crosstalk, xiao2016full}, which arise from multiply scattered photons that travel to neighboring region when imaged onto an array detector while remaining coherent with the reference beam. As coherence and incoherence are the extremes of a continuum, we examine the effects of partial coherence as the general case. To do so, we model a partially coherent source as a 2D source with a non-zero lateral extent, consisting of a continuous distribution of point sources that are mutually incoherent, with an intensity distribution, $I_s(x,y)$. If this extended source is placed at the focal plane of collimating lens, which does not affect the coherence properties of the source, all the points will be collimated into non-interfering plane waves, propagating to the sample at angles given by the positions of the corresponding point sources.

In the coherent limit of this partially spatially coherent source where the extended source is a single point source ($I_s(x,y)\propto\delta(x,y)$), after the collimating lens, we obtain a single perfectly coherent plane wave, and we recover FF-OCT with coherent illumination. In the incoherent limit of an infinitely wide source ($I_s(x,y)\propto 1$), we obtain a superposition of many mutually incoherent plane waves with continuous angular coverage allowed by the collimating lens. Essentially, each mutually incoherent point source can be thought of as an independent channel or mode across which FF-OCT at a particular illumination angle is performed, where the larger the source, the more channels and therefore the wider angular range. Thus, FF-OCT with partially spatially coherent illumination has a TF similar to that of point-scanning OCT \cite{marks2009partially, sentenac2018unified} (lower right panel of Fig. \ref{fig:four_TFs_high_NA}), which obtains illumination angular diversity through a focused beam (Fig. \ref{fig:resolution_doubling} compares three ways of attaining this TF discussed in this paper). Thus, Eq. \ref{TF_OCT_point_scan} can be used to compute the TF of FF-OCT with partially coherent light, adjusting the angular spectrum $E(k_x,k_y)$ according to the incoherent source extent. 

\subsection{Transillumination OCT} \label{TOCT}
Instead of reusing the illumination path as the detection path, transillumination OCT features a separate detection path on the other side of the sample \cite{hee1993femtosecond, thomas2008fourier, wang2010high, van2018high}. The detection channel is typically 180\textdegree-opposite so that all of the light is collected in the absence of a sample (i.e., a bright-field configuration). Transillumination OCT is almost always implemented with focused illumination with point scanning rather than with unfocused illumination with full-field detection, the reason for which becomes clear when we analyze their TFs using the FDT. In fact, for each wavenumber within the broadband OCT source, the TF is the same as that of transmission DT with equal illumination and detection NAs, which is a low-pass, lateral-resolution-enhanced, doughnut/toroid-shaped filter (Figs. \ref{fig:TF_PSF_overview} and \ref{fig:transillumination}a). This is because both have transmissive imaging geometries with angular illumination diversity, achieved either sequentially via plane wave angle sweeping or simultaneously via focused illumination (i.e., Fig. \ref{fig:resolution_doubling}, except in transmission). As such, transillumination OCT with focused illumination is sensitive not only to the average RI of the sample (i.e., the DC component of the RI or scattering potential), but also to low-frequency RI variation and therefore, like transmission DT, has some axial resolvability, albeit limited. However, without focused illumination, the axial resolution is worse (cf. Fig. \ref{fig:TF_PSF_overview}, first vs. second columns). Transillumination OCT is thus perhaps an exception among techniques with ``OCT'' in its name, as the only one with a TF centered at the $\mathbf{k}=(0,0,0)$ rather than $\mathbf{k}=(0,0,2k_0)$. As such, elsewhere in this document, we will assume, when referring to an OCT TF, that it is a band-pass without explicitly clarifying that it is in reflection-mode.

For transillumination OCT, each wavenumber accesses almost the same information about the sample, with the equivalent DT TF isotropically scaled in $k$-space in proportion to the wavenumber, $k_0$. At first glance, it would appear that the wavenumber diversity does not add much benefit in terms of TF volume in 3D $k$-space, especially for FF-OCT with coherent illumination, as all the Ewald spheres are tangent to each other at the $k$-space origin \cite{sentenac2018unified}. While this may be true in the first Born approximation, in the presence of multiple scattering, wavenumber diversity has proven to be useful to discriminate the weakly or singly scattered fields (i.e., those that do obey the first Born approximation, corresponding to ballistic photons) from the multiply scattered fields, the latter of which will have propagated over a longer distance than the former \cite{hee1993femtosecond}. Furthermore, focused illumination with a confocal pinhole offers not only lateral resolution enhancement, but also an additional mechanism for spatially filtering multiply scattered light. This confocal gate is absent in the hypothetical full-field transillumination OCT, which would thus suffer from the same cross-talk issue as in reflective FF-OCT \cite{karamata2004spatially, marks2009partially, dhalla2010crosstalk, xiao2016full}. However, full-field transillumination OCT can theoretically achieve the same confocality with a spatially incoherent source or with sequential multiangle illumination, just as in reflective FF-OCT (Fig. \ref{fig:resolution_doubling}); however, these strategies, to our knowledge, have not been reported and would be interesting areas of future investigation.

\subsection{Multi-angle transillumination OCT}
Since transillumination OCT has anisotropic resolution, with better lateral than axial resolution, recently researchers have incorporated angular diversity to obtain isotropic resolution, limited by the original lateral resolution. Over a decade ago, proofs of concept were demonstrated on phantom samples \cite{thomas2008fourier, wang2010high}, and a similar reflection-mode concept was demonstrated even earlier \cite{zysk2003projected}. However, it wasn't until recently that it was demonstrated on optically thick biological samples as a technique referred to as optical coherence projection tomography (OCPT) \cite{van2018high}. In analogy with X-ray CT or projection tomography, multi-angle transillumination OCT employs relative rotation between the illumination/detection path pairs and the sample (though all existing approaches have employed pure sample rotation). The resulting sinograms are then used in the usual backprojection algorithm to generate the isotropic-resolution reconstruction. 

\begin{figure}
    \centering
    \includegraphics[width=.6\textwidth]{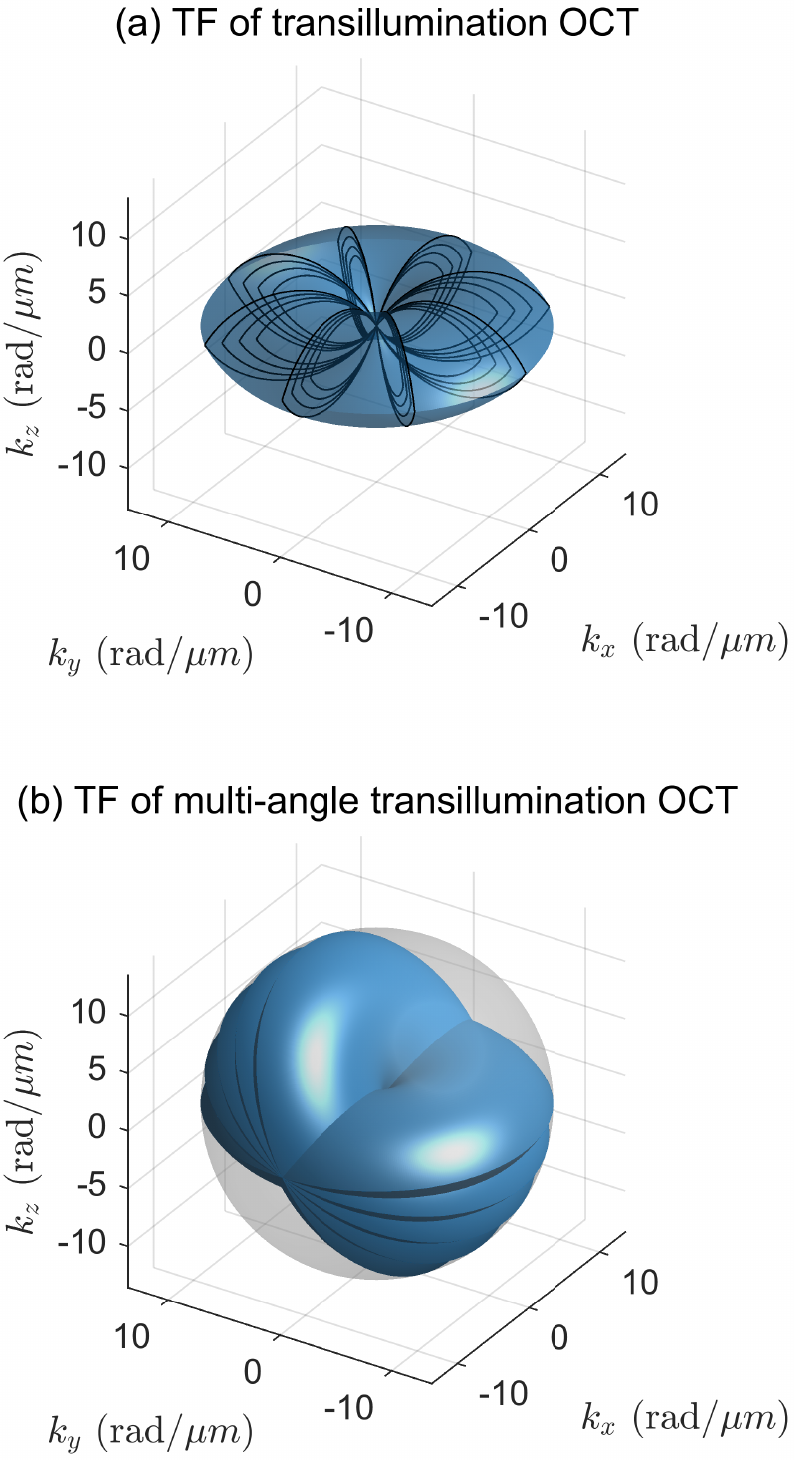}
    \caption{(a) TF of transillumination OCT, which is similar to that of transmission confocal microscopy and DT. The different rescaled versions of the cross-sections correspond to different wavenumbers within the broadband source. (b) TF of multi-angle transillumination OCT, which is obtained by rotating (a) about the $k_x$- or $k_z$-axis. The limiting surface is a sphere with a radius of $2k_{max}NA$.}
    \label{fig:transillumination}
\end{figure}

The key benefit of using transillumination OCT (with confocal detection) over monochromatic transmission confocal microscopy or DT, all three of which theoretically have similar TFs, is the ability to reject mulitply scattered fields, which are much more difficult to model. In particular, for each angle, as mentioned in the earlier section, the transillumination OCT allows isolation of the earliest-arriving photons, corresponding to singly scattered fields (assuming there is enough SNR to detect them, which depends on the source brightness and optical thickness of the sample), which obey the first Born approximation and whose TF can thus be modeled in our $k$-space theory as the doughnut/toroid-shaped low-pass structure. Thus, with data acquisition over 180\textdegree or 360\textdegree, the synthesized TF is a spherical ball with a cutoff radius of $2k_{max}NA$ (Fig. \ref{fig:transillumination}b). 

Note, however, the multiple scattering information is not completely discarded in multi-angle transillumination OCT. Specifically, in addition to scattering potential or RI information about the sample, transillumination OCT also allows measurement of integrated attenuation -- while the former is encoded in the time-of-flight of the ballistic peak, the latter is encoded in the amplitude of the peak. Thus, with multi-angle information, one can reconstruct not only an RI map, but also a map of the total attenuation coefficient, as was demonstrated in OCPT \cite{van2018high}. The attenuation, however, can be due to both scattering and absorption. 

\subsection{Effects of multiple scattering on transfer functions} \label{multiple_scattering}
Clearly, OCT does not cease to be useful when the first Born approximation is no longer satisfied. As Fig. \ref{fig:FDT} suggests, the effects of multiple scattering become significant when the field incident at a structure deeper within the sample is no longer accurately approximated by the incident illumination. Effectively, this means that multiple scattering simply aberrates or otherwise distorts the incident field at greater sample depths, such that the first Born approximation may be considered slab-wise valid \cite{chen2019multi} (although not jointly valid). Theoretically, we could do a plane-wave decomposition of this aberrated field, as we did for point-scanning OCT (Sec. \ref{point_scanning_OCT}), confocal microscopy (Sec. \ref{confocal}) and LF-OCT (Sec. \ref{line_field}), and thus as long as the beam maintains a forward-scattering bias, such that the angular spectrum is concentrated around DC, the TFs previously derived still approximately hold. As the multiple scattering becomes more severe to the point that the beam becomes completely random, with no memory of the original illumination, the angular spectrum is spread out with random phase, such that the multi-angle TFs in Eq. \ref{TF_OCT_point_scan} would combine in a phase-unstable manner (much like how phase instabilities across lateral scan positions for ISAM prevent a faithful depth-invariant resolution reconstruction).

\subsection{$k$-space interpretation of coherence gating} \label{coherence_gating}
Coherence gating is the ability to discriminate different scattering trajectories based on their optical path lengths using a broadband or low-temporal-coherence source. It is perhaps most intuitive to think about coherence gating in TD-OCT, in terms of photons propagating through and reflecting off of layers of a discrete, multi-layer structure. Under this model, interference between a reflection from a particular layer in the sample and reference beam only occurs if their path lengths are matched, with all other reflections ``gated'' out due to lack of temporal coherence with the reference beam. While this simplified picture is relatively straightforward to understand, we now generalize and interpret coherence gating in $k$-space in arbitrarily spatially inhomogeneous media. 

OCT is only sensitive to certain distributions of scattering potential or RI -- those with spatial frequency content lying in the OCT TF. In particular, discrete boundaries in the sample are discontinuities in the RI distribution of the sample that have spatial frequency content everywhere in $k$-space. Consider the following examples of types of discontinuities and their Fourier transforms (in 2D and neglecting some scale factors for simplicity):
\begin{enumerate}
    \item Point reflector:
    \begin{equation} \label{point_reflector}
        \delta(x-x_0,z-z_0)\overset{\mathcal{F}}{\leftrightarrow}\exp(-j(k_xx_0+k_zz_0)).
    \end{equation}
    \item Horizontal RI boundary:
    \begin{equation} \label{horizontal_boundary}
    H(z-z_0)\overset{\mathcal{F}}{\leftrightarrow}\left(\frac{1}{\pi k_z}+\delta(k_z)\right)\exp(-jk_zz_0),    
    \end{equation}
    where $H$ is the Heaviside step function.
    \item Small object:
    \begin{enumerate}[label=(\alph*)]
        \item 1D rectangular object:
        \begin{equation} \label{rect}
            \text{rect}((z-z_0)/w  )\overset{\mathcal{F}}{\leftrightarrow}\text{sinc}(wk_z)\exp(-jk_zz_0).
        \end{equation}
        \item Circular (or spherical) object:
        \begin{equation} \label{circ}
        \text{circ}((x-x_0)/w,(z-z_0)/w)\overset{\mathcal{F}}{\leftrightarrow}\text{jinc}\left(w\sqrt{k_x^2+k_z^2}
    \right)\exp(-j(k_xx_0+k_zz_0)),    
        \end{equation}
        where $\text{jinc}(x)=J_1(x)/x$ and $J_\alpha$ is the Bessel function of the first kind. 
    \end{enumerate}
\end{enumerate}
An actual biological sample may be a superposition of these examples, modeling RI discontinuities at, for example, cell or organelle boundaries.
Note that within an OCT TF, a small band-pass centered at $\mathbf{k}=(0,0,2k_0)$, all of these examples appear as sinusoidal fringes in $k$-space. This is obvious for a point reflector (Eq. \ref{point_reflector}). For a horizontal RI boundary (Eq. \ref{horizontal_boundary}), while the $k$-space response is a sinusoid with a decaying amplitude, within a small neighborhood surrounding $k_z=2k_0$, we have that
\begin{equation}
    \frac{\exp(-jk_zz_0)}{k_z}\approx \frac{\exp(-jk_zz_0)}{2k_0}.
\end{equation}
For a rect object (Eq. \ref{rect}), we can make the same type of approximation, so that 
\begin{equation}
\begin{split}
    \text{sinc}(wk_z)\exp(-jk_zz_0) \approx \frac{\sin(wk_z)}{2wk_0}\exp(-jk_zz_0)\\
    =\frac{j}{4wk_0}\left(\exp(jk_z(-z_0-w)-\exp(jk_z(-z_0+w)\right).
\end{split}
\end{equation}
In other words, we have two fringes, one with frequency $z_0-w$, the other $z_0+w$, corresponding to the reflections of the front and back boundaries of the rect function. The same argument can be made for a circular object (Eq. \ref{circ}), if we approximate the jinc function with a decaying sine. On the other hand, constant or slow-varying components of the RI distribution are not detected by reflection-mode OCT as they don't produce $k$-space sinusoids that intersect with its TF. These sinusoids are analogous to those we see in conventional 1D FD-OCT processing; however, in our $k$-space analysis, the sinusoids permeate throughout 3D $k$-space. 

We can thus interpret coherence gating in $k$-space as the orthogonality of 3D fringes of different frequencies. That is, if one takes the inner product of the Fourier transform of the scattering potential with a desired fringe, $\exp(-j(k_xx_0+k_yy_0+k_zz_0))$, across 3D $k$-space, only the fringe matching its frequency (i.e., 3D position), $(x_0, y_0, z_0)$, will produce a non-zero value, thus ``gating'' out the other frequencies/3D positions. This inner product is precisely the Fourier transform.

\section{Dispersion and aberrations}
Dispersion refers to the wavenumber-dependent changes in the refractive index of materials, such as glass, water, and biological tissue. If the amount of dispersion in the OCT sample and reference arms differs, the axial resolution of OCT images degrades \cite{Fercher2008, Izatt2015}. This axial degradation directly relates to the theory of ultra-short pulse broadening upon propagation through dispersive media (for a given source bandwidth, the temporal pulse width relates to the OCT axial resolution by $\delta z/c$). To compensate for dispersion mismatch, researchers either attempt to physically balance the amount of dispersion in both arms in hardware \cite{tearney1997high, drexler1999vivo, chen2004dispersion}, or digitally compensate by multiplying the interferogram by a phase factor, $\exp(j\phi(k))$, where $\phi(k)$ corresponds to the dispersion curve and is often expanded as a low-order polynomial \cite{fercher2001numerical, marks2003digital, wojtkowski2004ultrahigh,cense2004ultrahigh}.

While dispersion compensation typically corrects dispersion due to the imaging system (e.g., the lenses and optical fibers), in principle the sample being probed may also be a source of dispersion, which we discuss next. More generally, since we are presenting a full 3D $k$-space theory, we will also discuss the generalized 3D pupil, a generalization of dispersion, which corresponds primarily to the axial dimension, to include lateral ``dispersion,'' more commonly referred to as aberrations. In other words, just as the distinction between source spectrum vs. angular spectrum has been blurred in $k$-space theory, we can also put dispersion and aberrations on equal footing. Note that just as a distinction between imaging system-induced vs. sample-induced dispersion can be made, a similar distinction can be made between system-induced aberrations vs. sample-induced aberrations, where the former can analogously be accounted for through the angular spectrum of the input illumination.

\subsection{Effects of dispersion on OCT transfer functions} \label{sample_induced_dispersion}
While the scattering potential (Eq. \ref{scattering_potential}), the quantity of interest in the $k$-space framework, is defined for a single wavenumber, $k_0$, OCT uses multiple wavenumbers that could exhibit different RI and therefore scattering potentials in the sample of interest. Previous $k$-space-based analyses of OCT have also largely ignored this scattering potential dispersion, in part because sample-induced dispersion in most biological tissue samples is negligible, except for the largest bandwidths used in submicrometer-axial-resolution OCT \cite{povazay2002submicrometer,bizheva2017sub}. 

To examine the effects of scattering potential dispersion, analogous to how dispersion is conventionally handled to account for imaging system-induced dispersion (i.e., not sample-induced) \cite{Fercher2008}, we can perform a series expansion of the scattering potential. To make this analysis more tractable, we assume separability between the spatial and frequency dependence of the scattering potential,
\begin{equation}  \label{separable}
    V(\mathbf{r},k)=V(\mathbf{r})\exp(j\phi(k)),
\end{equation}
where $\phi(k)=0$ for a dispersion-less medium and $V(\mathbf{r})$ corresponds to Eq. \ref{scattering_potential}. The assumption behind the separability means that the dispersion curve does not depend on the spatial location (axially and laterally), which is a reasonable approximation when imaging biological tissue samples, which are largely composed of water. In fact, this assumption is often made in OCT, as, typically, global dispersion compensation values are used to correct for system dispersion, and sample dependent dispersion is ignored. There are a few works, however, that correct for axially or laterally dependent dispersion \cite{fercher2001numerical,lippok2012dispersion,pan2017depth,kho2019compensating}.

We can thus expand $\phi(k)$,
\begin{equation} \label{V_dispersion}
    \phi(k)=
    \phi(k_0)+
    \frac{d\phi}{dk}\Bigr\rvert_{k_0}(k-k_0)+
    \frac{1}{2}\frac{d^2\phi}{dk^2}\Bigr\rvert_{k_0}(k-k_0)^2+
    \frac{1}{6}\frac{d^3\phi}{dk^3}\Bigr\rvert_{k_0}(k-k_0)^3+
    ...\ .
\end{equation}
For a Fourier-domain OCT system, sampling $V(\mathbf{r},k)$ at various $k$, in principle we would use Eq. \ref{V_dispersion} as a normalization factor to extrapolate the value of $V(\mathbf{r},k=k_0)$. This procedure is analogous to digital dispersion compensation conventionally done in OCT to correct for system-induced dispersion \cite{Fercher2008}; thus, Eq. \ref{V_dispersion} compensates for both system-induced and sample-induced dispersion. A corollary of this observation is that as long as the first Born approximation is valid, we do not need to account for depth-dependent dispersion. Nevertheless, there are a few works that demonstrate techniques for depth-dependent dispersion compensation \cite{fercher2001numerical,lippok2012dispersion,pan2017depth,kho2019compensating}, implying that they were considering samples thicker or with more RI variation than supported by the first Born approximation.

\subsection{First-order dispersion}
While the second-order and higher-order terms are typically associated with axial resolution degradation, the zeroth and first-order terms are associated with axial resolution enhancement. To see this, assume a linear dispersion curve of the phase RI,
\begin{equation} \label{linear_dispersion}
    n(k)=n_0+Ck,
\end{equation}
where $C$ is constant with respect to $k$, with the corresponding group refractive index,
\begin{equation}
    n_g(k)=n(k)+k\frac{dn(k)}{dk}=n_0+2Ck.
\end{equation}
The group index is also frequently specified as a function of the vacuum wavelength (i.e., $n_g(\lambda)=n(\lambda)-\lambda dn(\lambda)/d\lambda$) or the frequency (i.e., $n_g(\omega)=n(\omega)+\omega dn(\omega)/d\omega$). Here, $k=\omega/c$ refers to the \textit{vacuum} wavenumber, as opposed to the medium wavenumber $k_m=nk$, so that the $n$ factor is observed to have a lengthening effect on the $k$-vectors,  therefore increasing the curvature of the Ewald spheres and having a resolution-enhancing effect. In particular, consider the medium $k$-bandwidth,
\begin{equation}
    \Delta k_m=n(k_2)k_2-n(k_1)k_1,
\end{equation}
where $k_1$ and $k_2$ define the extent of the source spectrum (e.g., FWHM). Substituting in Eq. \ref{linear_dispersion} and noting that $k_1+k_2=2k_0$, we obtain
\begin{equation}
    \Delta k_m = \Delta k(n_0+2Ck_0)=\Delta k n_g(k_0),
\end{equation}
which states that the vacuum bandwidth, $\Delta k$, is scaled by the medium's group index at the center wavenumber or wavelength. This is consistent with the well-known result that the OCT axial resolution inside the medium, as compared to that in air or a vacuum, is improved by a factor of the group index at the center wavelength. The general effect on the 3D OCT TF can be appreciated by considering the red and blue Ewald spheres in Fig. \ref{fig:FFOCT_TF_PSF}, whose radii are scaled by their respective phase RIs. 

\subsection{Generalized 3D pupil and numerical dispersion compensation}
Given that the $k$-space framework is agnostic to the distinction between the axial and lateral dimensions of the TF, similarly dispersion compensation in general should be considered jointly with 2D lateral spatial aberrations, a consideration which has been referred to as a generalization of numerical dispersion compensation \cite{adie2012guide,Adie2015}. Indeed, 2D pupil aberrations are commonly corrected numerically in Fourier ptychographic microscopy \cite{zheng2013wide, ou2014embedded, chung2019computational, konda2020fourier}, a coherent imaging technique that uses intensity-only images from multi-angle illumination to computationally reconstruct thin samples. In such applications, applying a 2D filter (i.e., multiplying by $\exp(j\phi(k_x, k_y))$) is sufficient, due to the thin-sample approximation. In OCT, however, instead of operating on the interferogram, which is a 1D function of wavenumber (i.e., multiplying by $\exp(j\phi(k))$), as is done in conventional numerical dispersion compensation, we can operate on the 3D $k$-space representation of the sample by multiplying by a combined phase filter that is a function of all three $k$-space coordinates (i.e., multiplying by $\exp(j\phi(k_x, k_y, k_z))$). Such a generalized phase filter may be regarded as a generalized 3D pupil \cite{adie2012computational, adie2012guide, Adie2015,mccutchen1964generalized} (where the ``pupil'' terminology originates from lateral aberration considerations), and is employed in computational adaptive optics (CAO) to correct for both monochromatic and chromatic aberrations \cite{adie2012computational,adie2012guide,Adie2015}. Although the weak scattering first Born approximation may limit the amount of such aberrations, these methods are still applicable in OCT, which in practice is often operated outside of the first Born approximation, or for correcting system-induced aberrations (e.g., astigmatism \cite{adie2012computational}). In weakly scattering or aberrating samples, the separability assumption in Eq. \ref{separable} may still be met, such that a 1D dispersion compensation phase factor and a 2D pupil function for aberrations may be employed independently, but in the general case the generalized 3D pupil may not be separable.

\label{generalized_dispersion}

\section{Limited depth of focus of OCT and computational refocusing with ISAM} \label{refocus}
One of the challenges of OCT is the trade-off between the lateral resolution and depth of focus. As a result, many practical OCT systems accept lateral resolutions on the order of 10 \textmu m or greater in order to obtain the hundreds-of-micron- to millimeter-scale depths of focus necessary for imaging practical samples. However, this seems to be at odds with the $k$-space framework, as the validity of TFs and PSFs requires shift invariance according to linear systems theory. Furthermore, we also established in Sec. \ref{holography} that since we are making complex field measurements, we should theoretically be able to digitally propagate the field to any plane, irrespective of the position of the focus.

\begin{figure}
    \centering
    \includegraphics[width=.8\columnwidth]{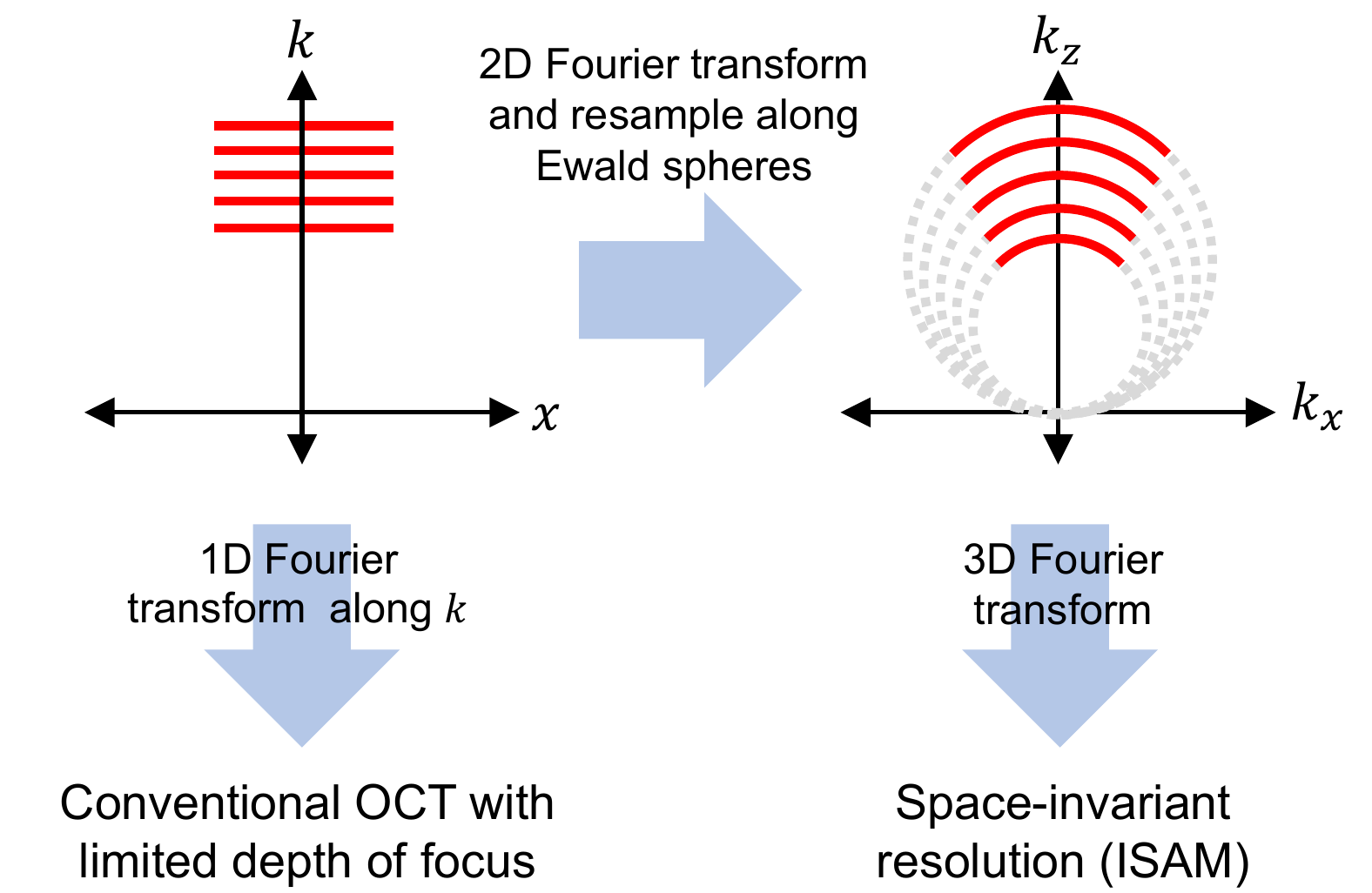}
    \caption{Conventional OCT processing produces images with limited depths of focus because they use 1D Fourier transforms along the wavenumber dimension. However, the measured information corresponds to non-planar manifolds in 3D $k$-space. For FF-OCT, these manifolds are the Ewald spheres, as depicted in this figure. Thus, to obtain space-invariant lateral resolution, one needs to resample in 3D $k$-space and perform a 3D Fourier transform.}
    \label{fig:resampling}
\end{figure}

\subsection{Importance of accounting for Ewald sphere curvature} \label{refocus_1}
The reason why OCT has limited depths of focus is that the typical OCT reconstruction algorithm in FD-OCT is a 1D inverse Fourier transform across the wavenumber sweep dimension, as the lateral dimensions are sampled in real-space (Fig. \ref{fig:resampling}). This is problematic because we have derived 3D TFs, which would require 3D inverse Fourier transforms for proper reconstructions. However, standard implementations of OCT acquire the lateral dimensions in the real space domain, and thus according to the FDT (Eq. \ref{FDT_eq}), we have the information corresponding to the 2D Fourier transform \textit{along the Ewald sphere} and not across the $k_xk_y$ plane, the ideal situation. Thus, the wavenumber sweep dimension does not correspond to $k_z$, but rather a distorted version thereof. Essentially, the culprit for the limited depth of focus of standard OCT is neglecting to account for the curvature of the Ewald sphere \cite{derosier2000correction}. In other words, standard OCT processing makes the erroneous assumption of separability of the lateral and axial TFs -- the higher the NA, the less valid this assumption, and thus the shorter the depth of focus. 

One procedure for correcting this distortion is to first take a 2D inverse Fourier transform across the two lateral dimensions, thus lifting the data back into 3D $k$-space (note that this step requires phase stability across lateral positions \cite{shemonski2014stability}). Next, for each wavenumber, we assign the information to the correct location in $k$-space along the Ewald sphere, in the case of FF-OCT (as we will see, the $k$-space surfaces for point-scanning and line-field OCT differ). In practice, resampling and interpolation are required. Finally, with all the information in the correct place in $k$-space, we take a 3D inverse Fourier transform to recover the depth-of-focus-independent resolution. In other words, the regions outside of the nominal depth of focus are refocused, at least to the extent allowed by the SNR.

In the following sub-sections, we will show that this correction procedure based on the FDT is an alternate derivation of inverse scattering theory for OCT, first derived almost 15 years ago and named by its inventors as interferometric synthetic aperture microscopy (ISAM) \cite{ralston2006inverse,ralston2007interferometric}. As will be seen, the resampling equation for FF-OCT differs from that of point-scanning OCT due to the difference in effective curvature of their respective TFs, as described above.

\begin{figure}
    \centering
    \includegraphics[width=.9\columnwidth]{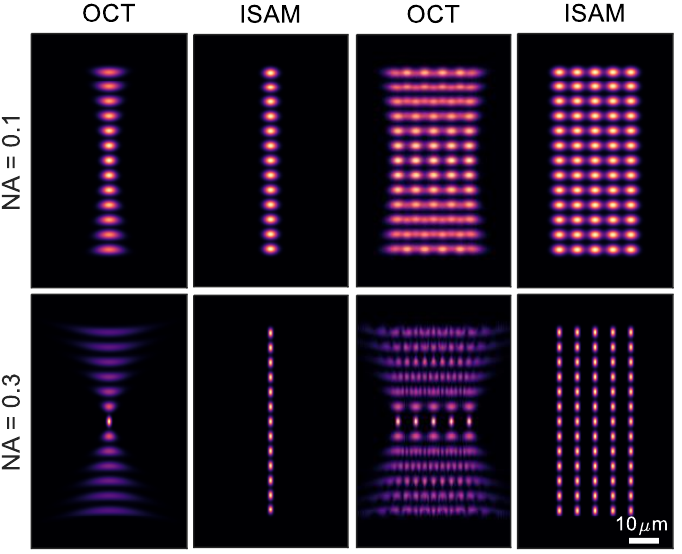}
    \caption{Resampling with interferometric synthetic aperture microscopy (ISAM) realizes depth-invariant lateral resolution. The panels show a comparison of OCT and ISAM for a single column of equally-spaced point scatterers (left two columns) and multiple columns of scatterers (right two columns) at NA=0.1 (top row) and NA=0.3 (bottom row), simulated using $\lambda_0=820$ nm and $\Delta\lambda=200$ nm. Note that away from the focus, the OCT response to the beads are curved and that the blurred spots do not superimpose incoherently.}
    \label{fig:ISAM}
\end{figure}

\subsection{Resampling for full-field OCT}
Although historically inverse scattering for FF-OCT \cite{marks2007inverse} (and a similar principle, holoscopy \cite{hillmann2011holoscopy,hillmann2012efficient}) postdates that for point-scanning OCT \cite{ralston2006inverse,ralston2006inverse_highNA,ralston2007interferometric}, we start with FF-OCT, whose resampling procedure is conceptually simpler as it operates directly with the Ewald sphere. We start with Eq. \ref{OCT_TF}, the correct form of the TF for FF-OCT, which we wish to distort into a form that allows separability between the axial and lateral components assumed in standard OCT processing. Once we find this distorting operation, we can invert it to obtain the correction procedure. We only need to focus on the argument of the first exponential factor in Eq. \ref{OCT_TF}, which contains a coupling between the axial and lateral components. Rewriting as
\begin{equation}
    \exp\left(\frac{-(k_r-2k_0\cos(k_\theta))^2}{8\sigma_k^2\cos^2(k_\theta)}\right)=\exp\left(\frac{-\left(\frac{k_r^2}{2k_z}-k_0\right)^2}{2\sigma_k^2}\right),
\end{equation}
we can see that a substitution $k_{\mathrm{\mathit{FFOCT}},z}=k_r^2/(2k_z)$ gives the separable form assumed in OCT, and that the lateral coordinates do not need to be modified, $k_{\mathrm{\mathit{FFOCT}},x}=k_x$ and $k_{\mathrm{\mathit{FFOCT}},y}=k_y$. Note that $k_{\mathrm{\mathit{FFOCT}},z}$ can be directly interpreted as the wavenumber sweep in swept-source OCT. Thus, the coordinate transformation is
\begin{equation} \label{FFOCT_resample}
\begin{split}
    &k_x=k_{\mathrm{\mathit{FFOCT}},x},\\
    &k_y=k_{\mathrm{\mathit{FFOCT}},y},\\
    &k_z=k_{\mathrm{\mathit{FFOCT}},z} \pm \sqrt{k_{\mathrm{\mathit{FFOCT}},z}^2-k_x^2-k_y^2}.
\end{split}
\end{equation}
Note that the two coordinate transformations for $k_z$ correspond to the top and bottom halves of the Ewald sphere (corresponding to reflective and transmissive geometries, respectively \cite{sheppard2012reconstruction}), centered at $(0, 0, k_{\mathrm{\mathit{FFOCT}},z})$ with radius $k_{\mathrm{\mathit{FFOCT}},z}$, and is consistent with a previous derivation by Marks et al. \cite{marks2007inverse}. Thus, every wavenumber in the sweep is corrected according to its respective Ewald sphere. This coordinate change becomes insignificant for small NAs in reflection ($k_z\rightarrow 2k_{\mathrm{\mathit{FFOCT}},z}$, where again the factor of 2 accounts for the round trip trajectory). Fig. \ref{fig:ISAM} compares reconstructions using standard OCT processing and using the coordinate resampling.

After the 3D $k$-space coordinate transform of Eq. \ref{FFOCT_resample} and interpolation of the scattering potential spectrum onto a regular grid for ease of digital processing (i.e., fast Fourier transform), we can take a 3D inverse Fourier transform to obtain a reconstruction with depth-independent resolution (Fig. \ref{fig:resampling}). For the sake of completeness, we should also multiply the volume element of the 3D inverse Fourier transform integral by the determinant of the Jacobian of coordinate change in Eq. \ref{FFOCT_resample}.

\subsection{Resampling for full-field OCT with off-axis illumination}
We can generalize the resampling equations in Eq. \ref{FFOCT_resample} by making the same argument for Eq. \ref{OCT_TF_general} as we did for Eq. \ref{OCT_TF}. We obtain a forward mapping of $k_{\mathrm{\mathit{FFOCT}},z}=k_r^2k_0/(2\mathbf{k}\cdot\mathbf{k_{illum}})$, from which after inverting, we obtain the following coordinate transformation:
\begin{equation} \label{FFOCT_resample_general}
\begin{split}
    &k_x=\frac{k_{\mathrm{\mathit{FFOCT}},z}k_{illum,x}}{k_0},\\
    &k_y=\frac{k_{\mathrm{\mathit{FFOCT}},z}k_{illum,y}}{k_0},\\
    &k_z=\frac{k_{\mathrm{\mathit{FFOCT}},z}k_{illum,z}}{k_0} \pm \sqrt{k_{\mathrm{\mathit{FFOCT}},z}^2-k_x^2-k_y^2},
\end{split}
\end{equation}
where we have taken advantage of the fact that the TF can be shifted to any position in 3D $k$-space while maintaining the real-space reconstruction up to a constant phase ramp (i.e., the Fourier shift theorem).
This coordinate resampling corresponds to an Ewald sphere centered at $\mathbf{k_{illum}}$ with radius $k_0$, and thus reduces to Eq. \ref{FFOCT_resample} for an on-axis illumination.

\subsection{Resampling for point-scanning OCT}
While the coordinate transformations for FF-OCT depend on the Ewald sphere, centered at $(0, 0, k_{illum})$ with radius $k_{illum}$, for point-scanning OCT \cite{ralston2006inverse,ralston2007interferometric}, they depend on a sphere centered at the origin with radius $2k_{illum}$. This is a consequence of Eq. \ref{TF_OCT_point_scan}, whose integral with respect to the input illumination direction is effectively revolving the Ewald sphere about the origin, thus creating a new spherical surface with twice the radius. Thus we can deduce the resampling equations to be
\begin{equation} \label{OCT_resample}
\begin{split}
    &k_x=k_{\mathrm{\mathit{OCT}},x},\\
    &k_y=k_{\mathrm{\mathit{OCT}},y},\\
    &k_z=\sqrt{4k_{\mathrm{\mathit{OCT}},z}^2-k_x^2-k_y^2},
\end{split}
\end{equation}
which are consistent with the derivation for ISAM for high NAs \cite{ralston2006inverse_highNA}.

However, we note that the information for a given wavenumber does not exclusively come from the spherical surface with radius $2k_{illum}$ \cite{sheppard2012reconstruction}, as can be easily visualized in the case of monochromatic confocal microscopy with a high NA (Fig. \ref{fig:four_TFs_high_NA}). Rather, the TF has high density along this sphere. Thus, alternative resampling curves may perform better, which can further be adjusted according to illuminations and pupils besides Gaussian \cite{sheppard2012reconstruction} (Sec. \ref{resampling_general}).

\subsection{Resampling for line-field OCT}

As mentioned in Sec. \ref{line_field}, the TF of LF-OCT has a horn torus shape for each wavenumber component. Thus, the coordinate transformation equations describe the partial surface of a horn torus facing away from the center:
\begin{equation} \label{LFOCT_resample}
\begin{split}
    &k_x=k_{\mathrm{\mathit{LFOCT}},x},\\
    &k_y=k_{\mathrm{\mathit{LFOCT}},y},\\
    &k_z=\sqrt{\left(k_{\mathrm{\mathit{LFOCT}},z}+\sqrt{k_{\mathrm{\mathit{LFOCT}},z}^2-k_y^2}\right)^2-k_x^2}.
\end{split}
\end{equation}
Note that along the $k_xk_z$-plane ($k_y=0$) and the $k_yk_z$-plane ($k_x=0$), this resampling scheme recapitulates the resampling scheme for point-scanning OCT (Eq. \ref{OCT_resample}) and FF-OCT (Eq. \ref{FFOCT_resample}), respectively (Fig. \ref{fig:LFOCT}). One interesting property of LF-OCT is that for a given wavenumber, there is more density along the toric surface than for point-scanning OCT along the spherical surface. This can be intuitively appreciated by the fact that the Ewald sphere shares its curvature with the torus along one lateral dimension. Similarly, in the case of FF-OCT, the Ewald sphere directly corresponds to the TF curvature in both lateral dimensions and thus all the density is concentrated along the Ewald sphere. This is, to our knowledge, the first derivation of ISAM resampling for LF-OCT.

\subsection{Resampling for other illumination patterns and pupils} \label{resampling_general}

Aside from plane waves or Gaussian beams focused in one or two dimensions, we may employ other types of illumination patterns, such as Bessel beams \cite{ding2002high, lee2008bessel, leitgeb2006extended, blatter2011extended} and annular pupils \cite{liu2011imaging, mo2013focus, yu2014depth}, as well as other detection strategies that differ from the illumination. Such experimental setups would result in TFs with different energy distributions and therefore different resampling relations. As noted in the preceding section, the resampling-based corrections are approximate because not all the $k$-space information for a given wavenumber resides along a 2D manifold, except for the case of FF-OCT for which that 2D manifold is the Ewald sphere \cite{sheppard2012reconstruction}. Thus, the resampling equations may have to be derived numerically by taking the center of mass through the TF for a given wavenumber. This was done using an OCT setup involving Bessel beam illumination and Gaussian mode detection \cite{coquoz2017interferometric}. The same reasoning would be applied to FF-OCT with partially coherent illumination, which has a similar TF to that of point-scanning OCT \cite{marks2009partially}.

A natural question that arises is whether there is an illumination or detection strategy such that no resampling is required -- that is, a TF whose energy distribution is maximized along a plane parallel to the $k_xk_y$-plane. This would be advantageous as it avoids the need for phase stability. One such strategy, as noted by Sheppard et al. \cite{sheppard2012reconstruction}, is to use annular pupils, which has been previously employed to extend the depth of focus of OCT \cite{yu2014depth}.

\section{Speckle in OCT}
Typically speckle in OCT is treated separately from its TF and modeled statistically \cite{schmitt1999speckle, bashkansky2000statistics, Fercher2008,dubose2017statistical, winetraub2019upper}. However, here, we show that speckle is a natural consequence of the band-pass nature of OCT TFs, as previously argued \cite{schmitt1999speckle}, and can be appreciated without invoking randomness and multiple scattering, which inevitably depend heavily on the sample structure. Our analysis will primarily focus on the OCT TF and make as few assumptions about the sample as possible. We start our analysis assuming separability of the axial and lateral TFs, which permits the use of Eqs. \ref{OCT_TF_approx}, \ref{axial_PSF}, and \ref{lateral_psf}, thereby simplifying the analysis. 

\subsection{Real-space interpretation of speckle}
\begin{figure}
    \centering
    \includegraphics[width=.75\columnwidth]{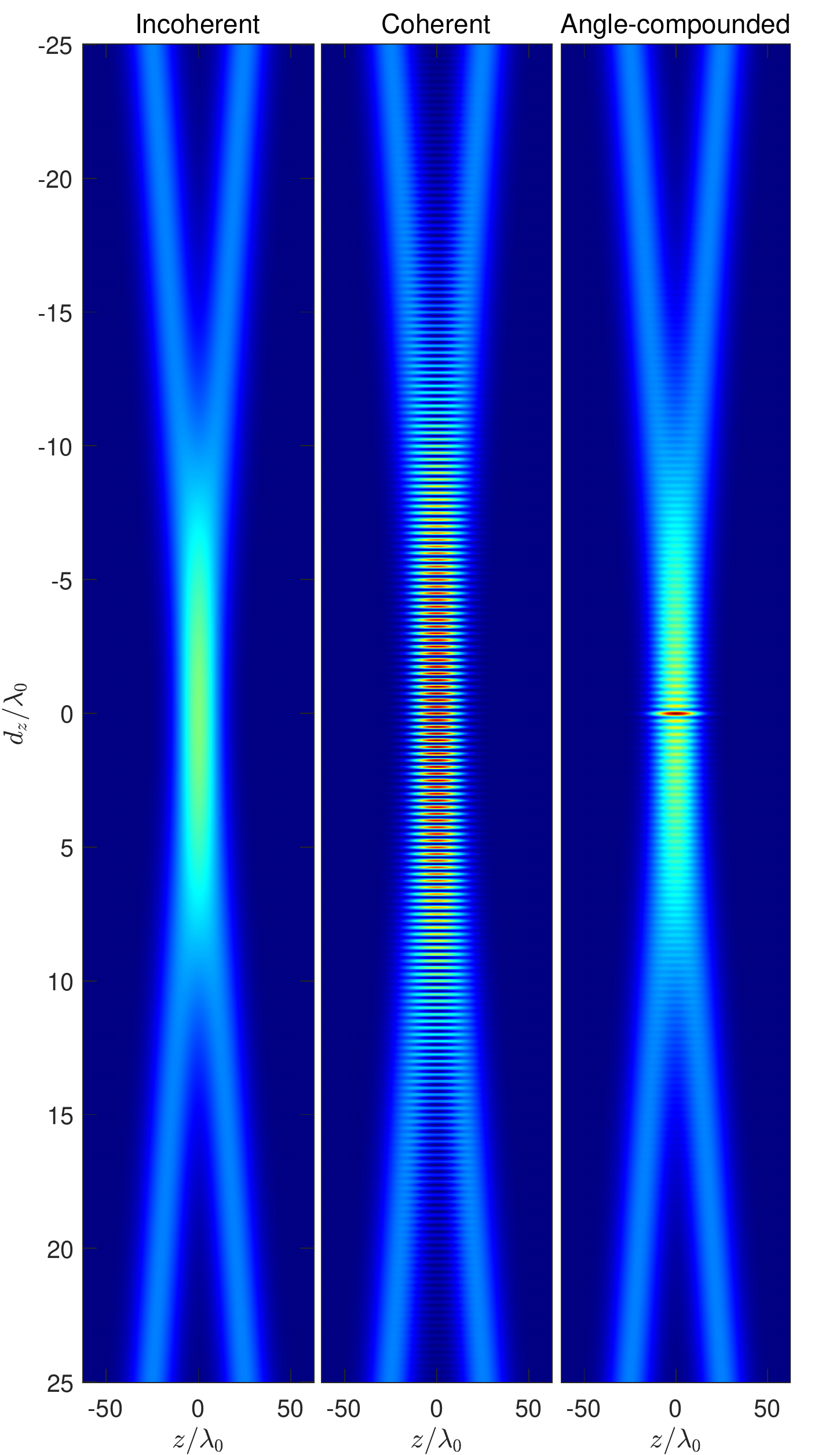}
    \caption{Angle compounding synthesizes incoherence and therefore reduces speckle. This figure compares responses of an incoherent model, coherent model, and an 180\degree\ angle compounded model to a pair of axially spaced scatterers (separation = $d_z$). The angle compounded result substantially reduces the coherent modulation artifacts, which would otherwise give rise to speckle. (Zoom into the figure to avoid aliasing)}
    \label{fig:speckle_angle_compounded}
\end{figure}

One common misconception about OCT is that its image formation is governed by convolution with an incoherent PSF. That is,
\begin{equation}
    \mathrm{\mathit{OCT}}(x,y,z)=|\mathrm{\mathit{psf}}(x,y,z)\otimes V(x,y,z)|^2\neq |\mathrm{\mathit{psf}}(x,y,z)|^2\otimes |V(x,y,z)|^2.
\end{equation}
Although OCT is based on low-coherence interferometry, OCT is very much a coherent imaging modality -- in our preceding $k$-space analyses, nowhere have we assumed incoherence or even partial coherence (except for FF-OCT in Sec. \ref{partial_coherence}). However, the coherent nature of the OCT TF is the very characteristic that confers speckle to OCT images. In other words, when the incoherent model agrees with the coherent model, that is tantamount to lack of speckle, as we will see. As speckle is caused by interference of sub-resolution scatterers, we start our analysis with the coherent PSF (Eqs. \ref{axial_PSF} and \ref{lateral_psf}) and analyze its effect on the simplest case of two scatterers, spaced by a potentially sub-resolution axial distance $d_z$ and lateral distance $d_x$ (omitting the $y$ dimension as it behaves in the same way as the $x$ dimension):
\begin{equation} \label{speckle_with_2_scatterers}
\begin{split}
    I(x,z)=&\left|\mathrm{\mathit{psf}}\left(x-\frac{d_x}{2},z-\frac{d_z}{2}\right)+\mathrm{\mathit{psf}}\left(x+\frac{d_x}{2},z+\frac{d_z}{2}\right)\right|^2 \\
    =&\left|\mathrm{\mathit{psf}}\left(x-\frac{d_x}{2},z-\frac{d_z}{2}\right)\right|^2+\left|\mathrm{\mathit{psf}}\left(x+\frac{d_x}{2},z+\frac{d_z}{2}\right)\right|^2 \\
    +&2\left|\mathrm{\mathit{psf}}\left(x-\frac{d_x}{2},z-\frac{d_z}{2}\right)\right|\left|\mathrm{\mathit{psf}}\left(x+\frac{d_x}{2},z+\frac{d_z}{2}\right)\right|\cos(2k_0d_z),
\end{split}
\end{equation}
where the first two terms in the expanded form correspond to those from a incoherent image formation model and the third term is the coherent interferometric term, which is the term of interest. Note that the oscillatory factor of the coherent term only depends on the axial separation, and not the lateral. When either the axial or lateral separation becomes large ($d_x,d_z\rightarrow\infty$), the coherent term goes to 0 and thus we obtain the incoherent image formation model. In other words, when there are no sub-resolution scatterers, the coherent and incoherent models agree with each other and there is no speckle. However, when the separation is comparable to the axial resolution, the coherent term induces high-contrast modulation, which gives rise to speckle. Fig. \ref{fig:speckle_angle_compounded} compares coherent to incoherent image formation models. Note that for certain sub-resolution separations, the two reflectors become nearly invisible due to destructive interference. 

While two scatterers may not give rise to speckle in the conventional sense, as the interference pattern is predictable and somewhat recognizable, when there are $N$ randomly-distributed scatterers within the system axial resolution we get a superposition of many pairwise coherent interferometric terms:
\begin{equation} \label{speckle}
\begin{split}
    I(x,z)=&\left|\sum_{n=1}^N r_n\mathrm{\mathit{psf}}(x-d_x^n,z-d_z^n) \right|^2\\
    =&
    \sum_{n=1}^N \left|r_n\mathrm{\mathit{psf}}(x-d_x^n,z-d_z^n)\right|^2\\+&
    \mathop{\sum^{N}\sum^{N}}_{n\neq m} 2r_nr_m
    |\mathrm{\mathit{psf}}(x-d_x^n,z-d_z^n)|
    |\mathrm{\mathit{psf}}(x-d_x^m,z-d_z^m)|
    \cos(2k_0(d_z^n-d_z^m)),
\end{split}
\end{equation}
where $\{r_n\}_{n=1}^N$ are arbitrary reflectivities and $\{d_x^n,d_z^n\}_{n=1}^N$ are the reflector positions. In this expression, as before, the first summation in the expanded form is the incoherent image formation model. The second double summation is the interferometric terms corresponding to each pair of reflectors in the resolution volume -- this is the speckle observed in OCT. In particular, when $\{d_x^n,d_z^n\}_{n=1}^N$ are randomly distributed, the double summation in Eq. \ref{speckle} is a superposition of randomly distributed fringes with random amplitudes. This treatment is consistent with the complex random walk interpretation of speckle, where here each $r_n\mathrm{\mathit{psf}}(x-d_x^n,z-d_z^n)$ is a complex phasor with a random phase and amplitude. The speckle is thus fully developed in the limit of a sum of a large number of phasors, which by the central limit theorem converges to a complex Gaussian distribution, thus giving rise to a Rayleigh distribution on the amplitudes, consistent with prior analysis of OCT speckle under the assumption of large collection of randomly distributed scatterers \cite{Fercher2008}. Eq. \ref{speckle_with_2_scatterers} may thus be regarded as a special case of underdeveloped speckle (perhaps the least developed speckle). In sum, analyzing the TF derived from $k$-space theory can explain speckle in OCT (at least that which occurs within the first Born approximation).

\subsection{\textit{k}-space interpretation of speckle}
\label{k_space_speckle}
We can reach similar conclusions about speckle in $k$-space. Once again, we consider the case of two closely spaced reflectors (separation, $d_z$, ignoring $d_x$ as it only affects the amplitude of the fringes), which manifest as beating of two fringes with very similar frequencies:
\begin{equation}
\begin{gathered} \label{speckle_k_space}
    \widetilde{A}(k)= 
    H_z(k)\left(\cos\left(2k\left(z-\frac{d_z}{2}\right)\right)+
    \cos\left(2k\left(z-\frac{d_z}{2}\right)\right)\right)=2H_z(k)\cos(kd_z)\cos(2kz),\\
    H_z(k)=\exp\left(\frac{-(k_z-2k_0)^2}{8\sigma_k^2}\right),
\end{gathered}
\end{equation}
where $H_z(k)$ is the axial component of the separable TF (Eq. \ref{OCT_TF_approx}), with DC terms not relevant for this analysis omitted. Eq. \ref{speckle_k_space} contains a high-frequency carrier corresponding to the average position of the two reflectors, modulated by a low-frequency envelope corresponding to their small separation, all of which is windowed by the TF. We argue that speckle arises when the period of the low-frequency beat envelope is comparable to or larger than the width of the TF; that is, when the separation $d_z$ is very small. Essentially, we can consider two extreme situations: 1) when the narrow TF is centered at the crest or trough of the low-frequency beat, or 2) when it is centered at a node of the beat. In the former case, we see only a single frequency at twice the amplitude, while in the latter case we see almost nothing. These correspond to, respectively, constructive and destructive interference. In other words, in both cases, the narrowbandedness of the OCT TF prevents us from seeing the bigger picture, that there are in fact two frequencies, not one. The larger the OCT TF (i.e., the larger the spectral and angular bandwidths), the better capable it is of detecting a low-frequency beat.

\subsection{Speckle reduction using angular compounding synthesizes incoherence} \label{speckle_reduction_section}
Since we have interpreted speckle as a consequence of the coherent nature of the TF and PSF, the motivation of speckle reduction is thus to make the image appear as if it was captured with an incoherent imaging system. While there are several methods for speckle reduction, here we focus on angular compounding, which involves incoherently averaging intensity OCT images acquired from multiple angles \cite{iftimia2003speckle, desjardins2006speckle, wang2009speckle}. Intuitively, angle-compounding speckle reduction works by changing the effective axial separation between the scatterers (i.e., modulating $d_z$), or equivalently changing the effective axial component of the illumination $k$-vector (i.e., modulating $k_0$), with the hope that the oscillating coherent terms in Eq. \ref{speckle_with_2_scatterers} and \ref{speckle} become averaged away to 0 or otherwise minimized. For this analysis, we focus on the case of two closely-spaced scatterers (axial separation=$d_z$), whose conclusions can be extended to the case of multiple scatterers, as we did in Eq. \ref{speckle}.
In particular, consider the OCT PSF (Eqs. \ref{axial_PSF} and \ref{lateral_psf}) rotated by angle, $\theta$, in the $xz$-plane,
\begin{equation}\label{rotated_psf}
\begin{split}
    \mathrm{\mathit{psf}}_\theta(x,z)=&
     \exp\left(-\frac{(x\cos(\theta) + z\sin(\theta))^2}{2\sigma_x^2}\right)
     \exp\left(-\frac{(x\sin(\theta) - z\cos(\theta))^2}{2\sigma_z^2}\right)\\
    &\times\exp\left(-j2k_0(x\sin(\theta)-z\cos(\theta)\right),
\end{split}
\end{equation}
where we have again omitted the $y$ dimension for simplicity. Then, the OCT response to the two scatterers separated by $d_z$ is given by
\begin{equation} \label{rotated_response_to_2_scatterers}
\begin{split}
    I_\theta(x=0,z=0)=& \left|\mathrm{\mathit{psf_\theta}}\left(0,\frac{d_z}{2}\right) +
    \mathrm{\mathit{psf_\theta}}\left(0,-\frac{d_z}{2}\right)
    \right|^2\\
    =& 2\exp\left(-\frac{d_z^2}{4}
    \left(\frac{\cos^2(\theta)}{\sigma_z^2}+
    \frac{\sin^2(\theta)}{\sigma_x^2}
    \right)\right)
    \left(1+\cos(2k_0d_z\cos(\theta))\right).
\end{split}
\end{equation}
While it is possible to analyze this equation at any arbitrary $xz$ position, the general expression is exceedingly complicated and distracts from the motivation to understand angle-compounded speckle reduction. Thus, here, we have set $x=0$ and $z=0$, which is the halfway point between the scatterers and is the position most affected by speckle, as the magnitude of the coherent term is maximized. From the coherent term in Eq. \ref{rotated_response_to_2_scatterers} (the cosine term), we can see that the effective axial separation for view angle $\theta$ is $d_z\cos(\theta)$ (an alternative interpretation is that the effective axial component of the illumination $k$-vector is $k_0\cos(\theta)$). Thus, angular compounding via incoherent superposition of the coherent term of Eq. \ref{rotated_response_to_2_scatterers} is given by

\begin{equation} \label{speckle_reduction}
\begin{split}
    S(d_z)=&\frac{1}{\pi}\int_{-\pi/2}^{\pi/2}
    \exp\left(-\frac{d_z^2}{4}
    \left(\frac{\cos^2(\theta)}{\sigma_z^2}+
    \frac{\sin^2(\theta)}{\sigma_x^2}
    \right)\right)
    \cos(2k_0d_z\cos(\theta))d\theta \\
    =& 
    \exp\left(-\frac{d_z^2}{4\sigma_z^2}\right)
    \frac{1}{\pi}
    \int_{-\pi/2}^{\pi/2}
    \exp\left(-\frac{d_z^2}{4}
    \left(\frac{1}{\sigma_x^2}-
    \frac{1}{\sigma_z^2}
    \right)\sin^2(\theta)
    \right)
    \cos(2k_0d_z\cos(\theta))d\theta \\
    \approx& 
    \exp\left(-\frac{d_z^2}{4\sigma_z^2}\right)
    \frac{1}{\pi}\int_{-\pi/2}^{\pi/2}
    \cos(2k_0d_z\cos(\theta))d\theta \\
    =&\exp\left(-\frac{d_z^2}{4\sigma_z^2}\right)J_0(2k_0d_z),
\end{split}
\end{equation}
where $J_\alpha$ is the Bessel function of the first kind.
We found that the approximation in Eq. \ref{speckle_reduction} is exact for isotropic resolution and holds well as long as $\sigma_z\not\ll\lambda_0$ or $\sigma_x\not\ll\lambda_0$, even for anisotropic PSFs. Intuitively, this can be understood by considering the fact that the Gaussian factor inside the integral of the second row of Eq. \ref{speckle_reduction} is broad and does not affect the rapidly oscillating cosine away from $\theta=0$, unless $\sigma_x$ or $\sigma_z$ is very small.

In sum, Eq. \ref{speckle_reduction} shows that angular compounding over a full 180\degree-range replaces the $\cos(2k_0d_z)$ in the original coherent term by $J_0(2k_0d_z)$, which decays like $1/\sqrt{2k_0d_z}$. Thus, with full angular compounding, scatterers can only contribute significantly to speckle over a length scale on the order of $1/k_0$ or a wavelength, compared to the typically much larger OCT axial resolution, $\sigma_z\propto1/\sigma_k$, given by the left factor in the fourth row of Eq. \ref{speckle_reduction}. While the coherent term, and therefore speckle, cannot be completely eliminated, even with full angular compounding, the term does become more delta-like. Therefore, the angularly compounded imaging system behaves like an incoherent imaging system over a larger domain (Fig. \ref{fig:speckle_angle_compounded}), while conventional OCT only does so when the scatterers are sparse, which is rarely the case in biological tissue. In this sense, angular compounding synthesizes incoherence. Note, however, that if the axial resolution of OCT is on the order of the wavelength \cite{povazay2002submicrometer,bizheva2017sub}, angular compounding is not as effective a strategy for speckle reduction, as the two factors in the fourth row of Eq. \ref{speckle_reduction} become comparable in width.

\begin{figure}
    \centering
    \centerline{\includegraphics[width=1.1\columnwidth]{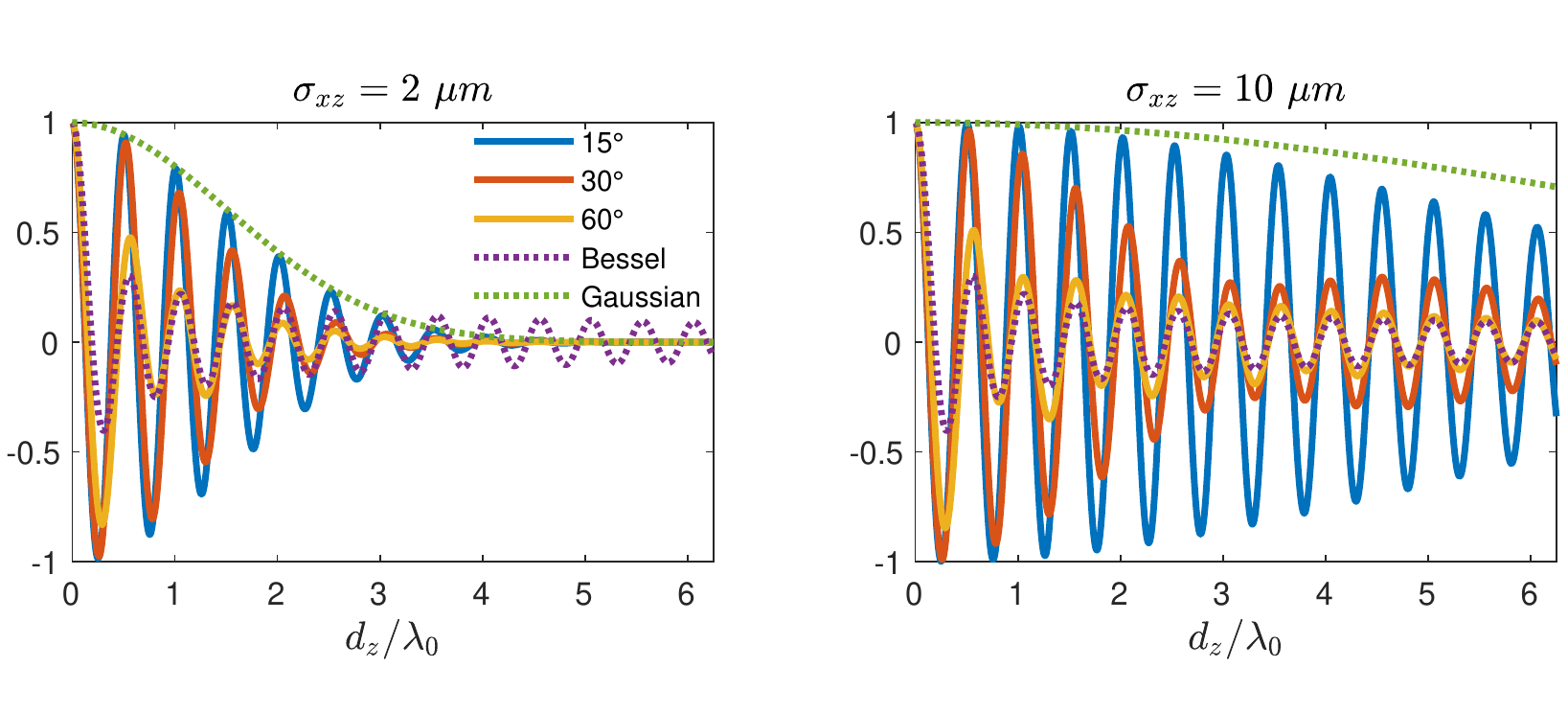}}
    \caption{Speckle reduction via angle compounding over a limited angular range ($\pm15\degree$, $\pm30\degree$, and $\pm60\degree$) with isotropic resolutions of 2 \textmu m and 10 \textmu m. For reference, the $J_0(2k_0d_z)$ (Bessel) and the $\exp(d_z^2/4/\sigma_z^2)$ (Gaussian) are plotted.}
    \label{fig:speckle_limited_angle}
\end{figure}

Finally, we consider angular compounding over a limited angular range, evaluating the first integral in Eq. \ref{speckle_reduction} (the unapproximated integral) numerically over an angular range of $\pm\theta_{max}$. These results are summarized in Fig. \ref{fig:speckle_limited_angle}, which show that limited angular compounding results in speckle reduction on length scales in between $\sigma_z$ and $\lambda_0$.

Increasing angular diversity is also a method of improving the resolution by expanding the TF of coherent imaging modalities. Later, we will discuss resolution enhancement techniques, including how they relate to speckle reduction (particularly the incoherent resolution enhancement techniques in Sec. \ref{incoherent_resolution_enhancement}).

\subsection{Equivalence of dynamic wavefront modulation and angular compounding for speckle reduction}
Another strategy for speckle reduction is to modulate the illumination wavefront with a dynamic, random phase mask (e.g., a translating diffuser). The idea is to present multiple uncorrelated phase masks during one integration period so as to average multiple speckle patterns. Such a strategy has been employed in point-scanning OCT \cite{liba2017speckle, zhang2019aperture} and in FF-OCT to reduce speckle due to lateral cross-talk \cite{povavzay2006full, stremplewski2019vivo} (albeit, from multiple scattering). Fundamentally, this approach has the same idea as angle-compounding-based speckle reduction, in that they both perform incoherent averaging over multiple independent coherent patterns. While angle-compounding-based speckle reduction typically involves incoherent digital averaging, wavefront-modulation-based speckle reduction achieves incoherent averaging by presenting multiple phase patterns at different times, thus preventing interference among these otherwise mutually coherent patterns. Thus, one could also collect multiple images with random illuminations and average them incoherently digitally; likewise, one could also sweep the illumination angle within one integration period. They differ in that they use different coherent bases, where angle-compounding-based approaches uses a multi-angle plane wave basis (or weakly focused waves in the case of point-scanning OCT), while wavefront-modulation-based approaches uses a random basis (each member of which is itself a coherent superposition of the former multi-angle basis). However, the end result is the same.

To appreciate the equivalence between angle compounding and wavefront modulation mathematically, we first model one particular modulated PSF as a coherent superposition of complex-valued PSFs from multiple angles, 
\begin{equation}
    \mathrm{\mathit{psf^{mod}_\phi}}(x,z)=\int_{\Theta} \mathrm{\mathit{psf_\theta}}(x,z)\exp(j\phi(\theta))d\theta,
\end{equation}
where integration is performed over some domain, $\Theta\in[\theta_{min}, \theta_{max}]$, restricted by the system NA, and $\mathrm{\mathit{psf_\theta}}$ is given by Eq. \ref{rotated_psf}, $\phi(\theta)$ is a random phase modulation introduced by the diffusing element at a particular time. Note that in general, the modulation can also include an amplitude component (i.e., $\phi(\theta)$ can be complex-valued). Let's once again analyze the case of two scatterers separated axially by $d_z$ (Sec. \ref{speckle_reduction_section}), with the same idea that our conclusions can be straightforwardly generalized to multiple scatterers. The OCT response for one modulation pattern, $\phi(\theta)$, is given by

\begin{equation} \label{OCT_modulated}
\begin{split}
    I_\phi(x=0,&z=0)\\
    &\begin{split}
    =\left| \mathrm{\mathit{psf^{mod}_\phi}}\left(0,\frac{d_z}{2}\right) +
    \mathrm{\mathit{psf^{mod}_\phi}}\left(0,-\frac{d_z}{2}\right)
    \right|^2
    \end{split}\\
    &\begin{split}
    =\left| 
    \int_\Theta  \left(
    \mathrm{\mathit{psf_\theta}}\left(0,\frac{d_z}{2}\right)+
    \mathrm{\mathit{psf_\theta}}\left(0,-\frac{d_z}{2}\right)\right)
    \exp(j\phi(\theta))
    d\theta
    \right|^2
    \end{split}\\
    &\begin{split}
        =\int_\Theta \int_\Theta
        \left(
        \mathrm{\mathit{psf_\alpha}}\left(0,\frac{d_z}{2}\right)+
        \mathrm{\mathit{psf_\alpha}}\left(0,-\frac{d_z}{2}\right)\right)
        &
        \left(
        \mathrm{\mathit{psf^{*}_\beta}}\left(0,\frac{d_z}{2}\right)+
        \mathrm{\mathit{psf^{*}_\beta}}\left(0,-\frac{d_z}{2}\right)\right)\\
        &\times\exp(j(\phi(\alpha)-\phi(\beta)))
        d\alpha d\beta
    \end{split}\\
    &
    \begin{split}
        \approx
        2\exp\left(-\frac{d_z^2}{4\sigma_z^2}\right)
        \int_\Theta \int_\Theta
        \big[&
        \cos(k_0d_z(\cos(\alpha)-\cos(\beta)))\\
        &+\cos(k_0d_z(\cos(\alpha)+\cos(\beta)))
        \big]
        \exp(j(\phi(\alpha)-\phi(\beta)))
        d\alpha d\beta,
    \end{split}
\end{split}
\end{equation}
where we have evaluated at $x=0$ and $z=0$ for the same reasons as in Eq. \ref{rotated_response_to_2_scatterers}, and expanded the square magnitude of the integral and substituted Eq. \ref{rotated_psf}. The approximation made here is similar to the one made in Eq. \ref{speckle_reduction}. The two cosine terms in the square brackets in Eq. \ref{OCT_modulated} are the incoherent, non-interfering terms ($\mathrm{\mathit{psf_\alpha}}\left(0,\frac{d_z}{2}\right) \mathrm{\mathit{psf^{*}_\beta}}\left(0,\frac{d_z}{2}\right) + \mathrm{\mathit{psf_\alpha}}\left(0,-\frac{d_z}{2}\right) \mathrm{\mathit{psf^{*}_\beta}}\left(0,-\frac{d_z}{2}\right)$) and the coherent, interferometric terms ($\mathrm{\mathit{psf_\alpha}}\left(0,\frac{d_z}{2}\right) \mathrm{\mathit{psf^{*}_\beta}}\left(0,-\frac{d_z}{2}\right) + \mathrm{\mathit{psf_\alpha}}\left(0,\frac{d_z}{2}\right) \mathrm{\mathit{psf^{*}_\beta}}\left(0,-\frac{d_z}{2}\right)$), respectively (the same distinction as we made in Eq. \ref{rotated_response_to_2_scatterers}). In the ensuing analysis, we will ignore the non-interferometric terms, only analyzing the interferometric terms, which are responsible for speckle (as we did in Eq. \ref{speckle_reduction}).

The OCT response upon \textit{incoherent} integration of the interferometric terms over multiple diffuser patterns is given by
\begin{equation} \label{wavefront_mod}
\begin{split}
    &\begin{split}
        S_{mod}(d_z)=2\exp\left(-\frac{d_z^2}{4\sigma_z^2}\right)
        \int_\Phi 
        \int_\Theta
        \int_\Theta
        &\cos(k_0d_z(\cos(\alpha)+\cos(\beta)))\\
        &\times\exp(j(\phi(\alpha)-\phi(\beta)))
        P(\phi)
        d\alpha d\beta
        d\phi
    \end{split}\\
    &\begin{split}
        \phantom{S_{mod}(d_z)}=2\exp\left(-\frac{d_z^2}{4\sigma_z^2}\right)
        \int_\Phi 
        \int_\Theta
        \int_\Theta
        \big[
            &\cos(k_0d_z\cos(\alpha))\cos(k_0d_z\cos(\beta))-\\
            &\sin(k_0d_z\cos(\alpha))\sin(k_0d_z\cos(\beta))
        \big]\\
        &\times\exp(j(\phi(\alpha)-\phi(\beta)))
        P(\phi)
        d\alpha d\beta
        d\phi,
    \end{split}
\end{split}
\end{equation}
where $\Phi$ is the domain of random phase modulation patterns accessible by the dynamic diffusing element, and $P(\phi)$ is the probability of a particular pattern (as a side note, the incoherent integration of the non-interferometric terms is the same as Eq. \ref{wavefront_mod}, except with the sign of the sine term in the square brackets reversed). Assuming that every modulation pattern is equally likely, so that $P(\phi)$ is constant and can be dropped from Eq. \ref{wavefront_mod}, and changing the order of integration, we have
\newcommand{\appropto}{\mathrel{\vcenter{
  \offinterlineskip\halign{\hfil$##$\cr
    \propto\cr\noalign{\kern2pt}\sim\cr\noalign{\kern-2pt}}}}}
\begin{equation} \label{S_mod}
\begin{split}
    \begin{split}
        S_{mod}(d_z)=2\exp\left(-\frac{d_z^2}{4\sigma_z^2}\right)
        \int_\Theta \int_\Theta
        \big[
            &\cos(k_0d_z\cos(\alpha))\cos(k_0d_z\cos(\beta))-\\
            &\sin(k_0d_z\cos(\alpha))\sin(k_0d_z\cos(\beta))
        \big]\\
        &\times\left[\int_\Phi 
        \exp(j(\phi(\alpha)-\phi(\beta)))
        d\phi\right]
        d\alpha d\beta.
    \end{split}
\end{split}
\end{equation}
Here, the factors contributing to wavefront modulation have been isolated in the square brackets, which is the mean outer product of the angularly-dependent modulation factors. We now consider two limiting cases that that result in different simplifications of this outer product: 1) $\phi$ is a constant, $\theta$-independent, deterministic phase, and 2) $\phi$ is random and follows an independent multivariate uniform distribution over $2\pi$ radians (i.e., $\phi(\theta)\sim \mathrm{\mathit{Unif}}(0, 2\pi)$). It's also possible for $\phi$ to have a stochastic and deterministic component (e.g., $\phi(\theta)\sim \mathrm{\mathit{Unif}}(0, \pi/4)$, which has a preferred phase), in which case the result would be a superposition of these two cases.

For case 1, the mean outer product in Eq. \ref{S_mod} is also constant and can be dropped. Thus, assuming that $\Theta\in[-\pi/2,\pi/2]$ to match the situation in Eq. \ref{speckle_reduction}, we have
\begin{equation}\label{deterministic}
\begin{split}
    &\begin{split}
    S_{mod}^{deterministic}(d_z)\propto
        2\exp\left(-\frac{d_z^2}{4\sigma_z^2}\right)
        \int_\Theta \int_\Theta
        \big[
            &\cos(k_0d_z\cos(\alpha))\cos(k_0d_z\cos(\beta))-\\
            &\sin(k_0d_z\cos(\alpha))\sin(k_0d_z\cos(\beta))
        \big]
        d\alpha d\beta
    \end{split}\\
    &\phantom{ S_{mod}^{deterministic}(d_z)}=
    2\exp\left(-\frac{d_z^2}{4\sigma_z^2}\right)
    \Bigg[
    \left(
    \int_\Theta
    \cos(k_0d_z\cos(\theta))
    d\theta
    \right)^2
    -
    \left(
    \int_\Theta
    \sin(k_0d_z\cos(\theta))
    d\theta
    \right)^2
    \Bigg]
    \\
    &\phantom{ S_{mod}^{deterministic}(d_z)}\propto
    2\exp\left(-\frac{d_z^2}{4\sigma_z^2}\right)
    \Big[
    J_0(k_0d_z)^2 -
    H_0(k_0d_z)^2
    \Big],
\end{split}
\end{equation}
where $H_\alpha$ is the Struve function. This case is simply just beam focusing, since all the multi-angle fields are mutually in phase and thus constructively interfere to form a focus. That is why Eq. \ref{deterministic} approaches 0 as $d_z$ approaches infinity, even in the absence of the Gaussian prefactor, as the focusing over a wide angular range improves the axial resolution.

For the more interesting case 2, the mean outer product in Eq. \ref{S_mod} converges to an integral over a delta function, because the integrand is 1 when $\alpha=\beta$ and otherwise a random value on the complex unit circle. Thus, integration over many random patterns will average away the off-diagonal components of the outer product to 0, leaving behind the identity matrix. Thus, assuming that $\Theta\in[-\pi/2,\pi/2]$ and $\Phi\in[-\pi/2,\pi/2]$ to match the situation in Eqs. \ref{speckle_reduction} and \ref{deterministic}, we have
\begin{equation}\label{speckle_reduction2}
\begin{split}
    &\begin{split}
        S_{mod}^{stochastic}(d_z)\appropto
        2\exp\left(-\frac{d_z^2}{4\sigma_z^2}\right)
        \int_\Theta \int_\Theta
        \big[
            &\cos(k_0d_z\cos(\alpha))\cos(k_0d_z\cos(\beta))-\\
            &\sin(k_0d_z\cos(\alpha))\sin(k_0d_z\cos(\beta))
        \big]\\
        &\times\left[\int_\Phi 
        \delta(\alpha-\phi)\delta(\beta-\phi)
        d\phi\right]
        d\alpha d\beta
    \end{split}\\
    &
    \begin{split}
        \phantom{ S_{mod}^{stochastic}(d_z)}=
        2\exp\left(-\frac{d_z^2}{4\sigma_z^2}\right)
        \int_\Phi
        \Bigg[
        &\left(\int_\Theta
        \cos(k_0d_z\cos(\theta))
        \delta(\theta-\phi)
        d\theta
        \right)^2\\
        -
        &\left(\int_\Theta
        \sin(k_0d_z\cos(\theta))
        \delta(\theta-\phi)
        d\theta
        \right)^2
        \Bigg]
        d\phi
    \end{split}\\
    &\phantom{ S_{mod}^{stochastic}(d_z)}=
    2\exp\left(-\frac{d_z^2}{4\sigma_z^2}\right)
    \int_\Phi
    \big[
    \cos^2(k_0d_z\cos(\phi))-
    \sin^2(k_0d_z\cos(\phi))
    \big]
    d\phi\\
    &\phantom{ S_{mod}^{stochastic}(d_z)}=
    2\exp\left(-\frac{d_z^2}{4\sigma_z^2}\right)
    \int_\Phi
    \cos(2k_0d_z\cos(\phi))
    d\phi\\
    &\phantom{ S_{mod}^{stochastic}(d_z)}
    \propto \exp\left(-\frac{d_z^2}{4\sigma_z^2}\right)J_0(2k_0d_z),
\end{split}
\end{equation}
where we used the sifting property of the delta function to evaluate the inner integrals. This result is identical to angle-compounding based speckle reduction result (Eq. \ref{speckle_reduction}). In fact, the first line in Eq. \ref{speckle_reduction2} can be interpreted as using a delta amplitude modulation function with a fixed phase that is swept across the full angular range, thereby modeling sweeping of the illumination angle during one incoherent integration period -- this is precisely angle-compounding-based speckle reduction. As a side note, the result for the non-interferometric term is the same, except with the sign of the sine term in Eq. \ref{speckle_reduction2} flipped, resulting only the Gaussian prefactor -- this is the same as for angle compounding, where the non-interferometric term is simply the Gaussian prefactor for all angles prior to compounding (see Eqs. \ref{rotated_response_to_2_scatterers} and \ref{speckle_reduction}).

\begin{figure}
    \centering
    \includegraphics[width=.85\textwidth]{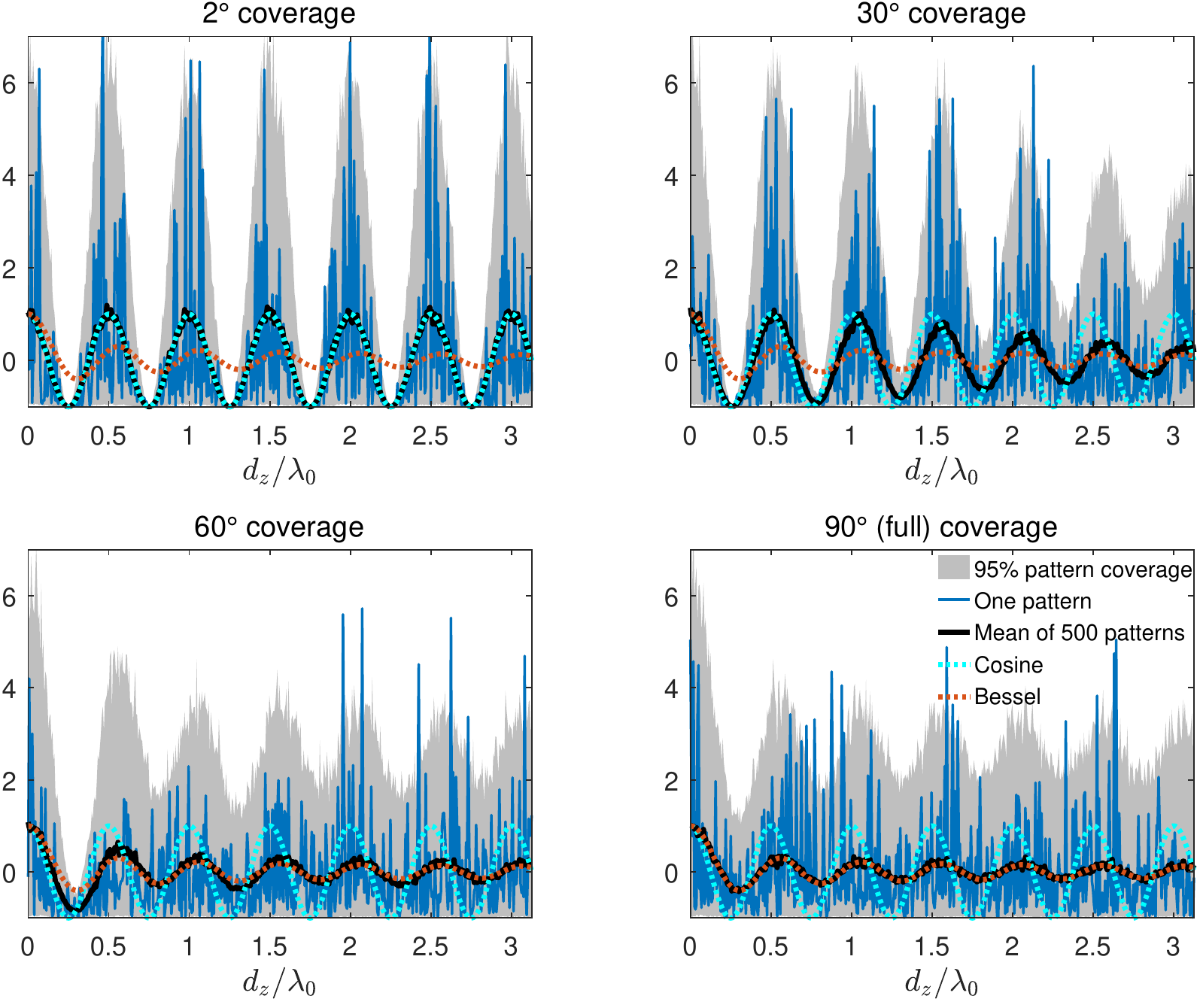}
    \caption{Simulation of incoherent averaging of 500 modulated wavefront patterns as a function of separation between the two scatterers for multiple angular ranges. The gray regions capture 95\% of the modulation patterns. The argument of the Bessel and cosine functions is $2k_0d_z$. This simulation ignores the Gaussian prefactor.}
    \label{fig:wavefront_mod}
\end{figure}

In sum, both angle compounding and wavefront modulation obtain the same degree of speckle reduction. Of course, these are theoretical results that rely on the ideal conditions of full angular coverage ($\pm\pi/2$) and incoherent integration over an infinite number of independent speckle patterns and angles, which we have chosen to facilitate analytical evaluation of integrals. In practice, the angular range will be limited by the NAs of practical objectives, and the number of angles or independent modulation patterns will be finite. We can, however, conclude that angular compounding and wavefront modulation over a limited angular range will asymptotically have the same speckle reduction performance, because Eqs. \ref{speckle_reduction} and \ref{speckle_reduction2} reduce to the same integral over angular range. Fig. \ref{fig:wavefront_mod} shows simulations of Eq. \ref{speckle_reduction2} using a finite number of modulation patterns over multiple angular ranges, without the Gaussian prefactor, which would otherwise suppress speckle at larger inter-scatterer separation (c.f., Fig. \ref{fig:speckle_limited_angle}). Furthermore, the modulation patterns may follow other distributions other than the uniform distribution, in which case $P(\phi)$ cannot be dropped from Eq. \ref{wavefront_mod}. In these cases, Eqs. \ref{OCT_modulated} and \ref{wavefront_mod} would likely have to be simulated numerically with discrete sums. 

Finally, we also reiterate that the conclusions drawn here for angle-compounding- and wavefront-modulation-based speckle reduction refer to speckle in the first Born or single scattering regime. Modeling reduction of speckle due to multiple scattering is more involved, as it precludes simplification to a broadly generalizable non-stochastic two-scatterer sample model and it relies more on the sample properties.

\section{Coherent resolution enhancement in OCT}
\label{coherent_resolution_enhancement}
Coherent resolution enhancement involves expanding the area of the TF in $k$-space, which can be achieved using a few strategies generally centered around the idea of increasing the angular or spectral bandwidths. Strategies referring to the expansion of the angular bandwidth are typically referred to as synthetic aperture techniques. These strategies require measurement diversity, such as through lateral scanning of a focused point (ISAM), angular scanning of a plane wave, or somewhere in between (lateral and angular scanning of a focused beam). Because synthetic aperture techniques must coherently combine information from multiple measurements, there needs to be a fixed or otherwise predictable phase coherence among the measurements (i.e., phase stability).

\subsection{Synthetic aperture techniques}

Perhaps the most conceptually straightforward approach to achieve high resolution would be to use larger-NA objectives and larger spectral bandwidths combined with ISAM or any of the resampling techniques described in Sec. \ref{refocus}, with which in theory one would be able to achieve the TFs derived in Sec. \ref{transfer_functions}. ISAM achieves angular diversity through the high-NA illumination, but must use lateral scanning so that every 3D spatial position within the sample observes the angular diversity. In particular, only the position corresponding to the nominal focus of the beam observes all of the plane wave angles in phase. Without lateral scanning, there are two effects: 1) neighboring positions in the lateral plane receive no light and therefore have zero angular diversity because of destructive interference, and 2) the further away axially from the focus, the more divergent the beam and the more locally plane-wave-like the beam, and therefore the less angular diversity (i.e., defocus) -- angular diversity is synthesized by translating the beam laterally, so that a given defocused position observes the diverging wavefront everywhere. This is the reason why ISAM is termed so -- angular coverage is synthesized away from the nominal focus, which without correction is the only position that observes angular diversity.

If we regard the maximum NA as 1.0 and assume that the refractive index part, $n$, of the NA expression serves to shorten the wavelength (or lengthen the wavenumber), the maximum $k_x$ or $k_y$ range for FF-OCT is $\pm nk_{max}$, where $k_{max}$ is the largest vacuum wavenumber used to illuminate the sample. Using a focused illumination with the same NA of 1.0, the maximum $k_x$ or $k_y$ range doubles, becoming $\pm 2nk_{max}$. 
As a result, the TFs are always constrained to reside inside a sphere of radius $2nk_{max}$.

This $2nk_{max}$-radius limit is the same as for diffraction tomography, which can achieve the same TF as the point-scanning analog through angular diversity of illumination through sequential plane wave illumination with wide-field detection rather than through focused illumination. In analogy to diffraction tomography and other synthetic aperture techniques such as synthetic aperture radar, thus another strategy for synthetic aperture in OCT is to acquire OCT images over a potentially smaller aperture, but alter the illumination angles sequentially to synthesize a larger TF for FF-OCT (e.g., Eq. \ref{OCT_TF_general}). One could either rotate the sample so that illumination and collection are along the same axis, or keep the aperture fixed and vary the illumination angle (Eq. \ref{OCT_TF_general}), producing distinct TFs. Yet a third strategy would be a combination of these two strategies, whereby a focused beam could be scanned laterally and rotated. However, in all cases the synthesized TF would still be constrained by the $2nk_{max}$-radius sphere. 

\subsection{Lateral point-scanning vs. angular plane wave rotation: a practical distinction in SNR distribution} \label{coherent_approaches}
Fundamentally, all of these strategies are the same if we consider the effective plane wave angular coverage, differing only by the order in which the angularly-varying plane waves are delivered to the sample. However, there are at least two important practical differences among these strategies: SNR and phase stability. 

SNR considerations stem from the fact that even though our $k$-space analyses uses optical fields (i.e., in the FDT, Eq. \ref{FDT_eq}), in practice we can only detect intensity, which is proportional to the magnitude of the field squared. This is important because measurement SNR is related to the number of photons detected, which is proportional to the intensity rather than the field amplitude. One way to think of this is that there is a fixed energy (photon) budget, with which we are free to distribute spatially across the sample via constructive and destructive interference of our multi-angle plane waves by adjusting their phase and amplitudes (i.e., the angular spectrum). For example, the simplest case might be a single plane wave, which allocates our energy equally across the sample (in the absence of multiple scattering, as afforded by the first Born approximation), thereby conferring more uniform SNR across the sample. We can also choose the phase and amplitudes of our multi-angle plane waves such that they form a tightly-focused Gaussian beam -- in this case, most of our energy budget is allocated to a region surrounding the focus, meaning there is high SNR at and near the focus, but low SNR everywhere else. Because SNR is low elsewhere, scanning is required. This is a major limitation of ISAM with high-NA illumination -- although the lateral resolution theoretically becomes depth-independent, the SNR is only appreciable within the depth of focus. Contrast this with sweeping the plane wave angle sequentially as is done in diffraction tomography (and potentially in FF-OCT via Eq. \ref{OCT_TF_general}), which achieves the same depth-independent resolution as ISAM, but with a more spatially uniform SNR (also meaning lower SNR at the focus than in ISAM). Other strategies are possible, such as intentionally introducing astigmatism into the beam \cite{adie2012computational} or using a Bessel beam \cite{ding2002high, lee2008bessel, leitgeb2006extended, blatter2011extended}, which achieve moderate SNR over a longer depth range. It remains to be seen whether there is an illumination strategy that confers wide angular diversity at every spatial position of the sample, thus reducing or altogether obviating the need for scanning of any kind.

The other practical consideration is phase stability among the sequential measurements \cite{shemonski2014stability}, which in practice often means nanometer-scale motion stability. For ISAM, there needs to be phase stability as the beam is translated laterally, which may be compromised due to sample motion or jitter in the scanning mechanism (e.g., galvanometers). Translational phase stability allows us to take 2D Fourier transforms across the lateral dimensions in order to operate in 3D $k$-space (Sec. \ref{refocus_1}). For FF-OCT, the lateral components of the 2D backscattered field are detected simultaneously, thus conferring lateral phase stability \cite{leitgeb2019face}. However, sample motion is still a concern due to the slower source sweep rate or mechanical axial translation, which can affect the axial phase stability \cite{hendargo2011doppler, mece2020high}. Furthermore, phase stability must be maintained across different illumination angles, so that the Ewald spheres can be constructively superimposed in 3D $k$-space.

\section{Incoherent resolution enhancement in OCT}
\label{incoherent_resolution_enhancement}
Finally, we discuss incoherent resolution enhancement techniques, a recent development that does not attempt to reconcile the phase relationships among the sequential measurements. To clarify, we are not referring to super-resolution in incoherent imaging techniques like fluorescence microscopy, but rather we are focusing on enhancing resolution in coherent imaging techniques such as OCT using techniques that do not rely on phase information. 

\subsection{Optical coherence refraction tomography (OCRT)}
\begin{figure}
    \centering
    \includegraphics[width=.75\columnwidth]{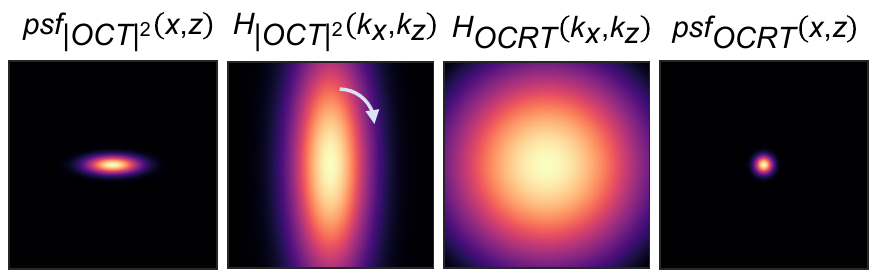}
    \caption{Optical coherence refraction tomography (OCRT) obtains resolution enhancement and speckle reduction. OCRT uses intensity OCT images, which have anisotropic PSFs and TFs, from multiple angles to reconstruct an image with isotropic resolution, limited by the original OCT axial resolution.}
    \label{fig:OCRT}
\end{figure}

One technique that achieves incoherent resolution enhancement in OCT is one we recently introduced and named optical coherence refraction tomography (OCRT) \cite{zhou2019optical,zhou2020spectroscopic}. One motivation of OCRT is that OCT typically has anisotropic resolution, with the axial better than the lateral resolution, which is due to the desire to have long depths of focus on the order of hundreds of microns to millimeters for bulk tissue imaging. To combat this anisotropy, OCRT uses magnitude OCT images (i.e., with phase information discarded) from multiple angles to create a reconstruction with more isotropic resolution, limited by the axial resolution (or lateral, whichever is better \cite{zhou2020spectroscopic}). The theory of OCRT was previously explained in analogy to X-ray computed tomography (CT) using an anisotropic TF centered at the origin of $k$-space \cite{zhou2019optical}. With sample rotation or angular steering of the illumination, the TF rotates and the superposition approaches an isotropic TF (Fig. \ref{fig:OCRT}). The requirement that the TFs be centered at the $k$-space origin appears to be at odds with the band-pass structure we derived for OCT and other reflective coherent imaging modalities (Sec. \ref{transfer_functions}). However, if we make certain assumptions about the sample, we will see that the DC-centered TFs are, in fact, consistent with our $k$-space framework, which we will now demonstrate.

For the ensuing analysis of OCRT, we are making the separability assumption of the OCT TF (Eq. \ref{OCT_TF_approx}), which is reasonable because without phase information and therefore ISAM resampling to rely on, in practice OCRT (and OCT in general) uses low NAs to obtain relatively uniform SNR and lateral resolution over long depths of focus. As a result of this separability assumption, we can drop the $y$ dimension as its analysis is identical to that of $x$.

Continuing, the key enabling assumption that justifies a DC-centered TF in OCT is that the sample is a finite set of randomly distributed discrete reflectors (i.e., a superposition of delta functions). This assumption is identical to the one made in incoherent speckle reduction discussed previously (Sec. \ref{speckle_reduction_section}), and often in OCT in general. Thus,
\begin{equation}
\label{sum_assumption}
    V(x,z)=\sum_{n=1}^N r_n\delta(x-x_n, z-z_n),
\end{equation}
with random amplitudes $\{r_n\}_{n=1}^N$ and positions $\{x_n,z_n\}_{n=1}^N$. Note that this is a reasonable assumption, even in biological tissue with slow RI variation in addition to scatterers, because the band-pass nature of the OCT TF cannot detect the low-frequency RI information anyway. Under this assumption, in $k$-space, the sample scattering potential is a superposition of complex exponentials with frequencies given by the spatial positions:
\begin{equation} \label{discrete_sum_k_space}
    \widetilde{V}(k_x,k_z)=\sum_{n=1}^N r_n\exp(-j(k_xx_n+k_zz_n)).
\end{equation}
Note that this scattering potential spectrum has infinite bandwidth, and thus in theory it does not matter which pass-band we choose, as each term in the sum exists everywhere, as ensured by the randomness of the $\{x_n,z_n\}_{n=1}^N$ \cite{Ralston2015}. In practice, this means we can use any source bandwidth and image the sample from any angle, and we would still be able to detect the sample scatterers. An example of a sample that violates our discrete-sum assumption is a pristine glass cover slide, a sample whose scattering potential is concentrated along a straight line in $k$-space. Thus, there are view angles at which OCT observes nothing (e.g., when the incident beam is not nearly orthogonal to the surface). However, for biological samples, the discrete-sum assumption is reasonable.

Proceeding with this assumption, we compute the $k$-space coverage of an intensity OCT image, with the phase discarded. First, in preparation for the next step, we note that 
\begin{equation}
\mathrm{\mathit{psf}}_\mathrm{\mathit{lpf}}(x,z) = 
\left|\mathrm{\mathit{psf}}(x,z) \right|^2\propto\exp\left(-\frac{x^2}{\sigma_{xy}^2}-\frac{z^2}{\sigma_z^2}\right)
\overset{\mathcal{F}}{\leftrightarrow}
\exp\left(-\frac{1}{4}\left(\sigma_{xy}^2k_x^2+\sigma_z^2k_z^2\right)\right)
\propto H_\mathrm{\mathit{lpf}}(k_x,k_z),
\end{equation}
which is the incoherent PSF and DC-centered TF that we seek ($\mathrm{\mathit{lpf}}$ = low-pass filter). We then compute the Fourier transform of Eq. \ref{speckle}, which is the square magnitude OCT response to a discrete sum of reflectors, and obtain
\begin{equation} \label{OCT_intensity}
\begin{split}
    \widetilde{I}(k_x,k_z)\propto 
    H_\mathrm{\mathit{lpf}}(k_x,k_z)\times 
    \left[\begin{split}
    &\phantom{+}\sum_{n=1}^N r_n^2 \exp(-j(k_x d_x^n + k_z d_z^n)) \\
    &+ \mathop{\sum^{N}\sum^{N}}_{n\neq m}
    \begin{aligned}[t]  
    &\Biggl[2r_nr_m\cos(2k_0(d_z^n-d_z^m)) \\
    &\times\exp\left(-j\frac{1}{2}\left(k_x(d_x^n+d_x^m)+k_z(d_z^n+d_z^m)\right)\right)\\
    &\times\exp\left(-\frac{(d_x^n-d_x^m)^2}{4\sigma_{xy}^2}-\frac{(d_z^n-d_z^m)^2}{4\sigma_z^2}\right)
    \Biggr]
    \end{aligned}
    \end{split}\right].
\end{split}
\end{equation}
Essentially, this is the autocorrelation of the OCT band-pass TF. This equation is the Fourier spectrum of the discrete-sum sample, filtered by the apparently low-pass, DC-centered TF, $H_\mathrm{\mathit{lpf}}(k_x,k_z)$. That is, the terms in the multi-line brackets can be interpreted as the sample, which we now analyze line by line. The first line (the single summation) is functionally identical to the assumed scattering potential spectrum of our discrete-sum sample (Eq. \ref{discrete_sum_k_space}), while the remaining terms (the double summation) can be attributed to speckle. The factors in the second and third row of the multi-line brackets contain the beat and carrier frequencies, respectively, due to two closely axially spaced reflectors (completely analogous to our $k$-space interpretation of speckle in Eq. \ref{speckle_k_space}). The factor in the fourth row of the multi-line brackets is a real-valued scaling factor that goes to 0 when the separation between the $n^{th}$ and $m^{th}$ reflectors becomes large compared to the resolution. This is also consistent with our previous analysis of speckle, as two reflectors only contribute significantly to speckle when the separations are sub-resolution. 

Now that we have demonstrated that the TF of the intensity OCT image is apparently centered at the $k$-space origin (i.e., $H_\mathrm{|\mathit{OCT}|^2}(k_x,k_z)=H_\mathrm{\mathit{lpf}}(k_x,k_z)$ and $\mathit{psf}_\mathrm{|\mathit{OCT}|^2}(x,z)=\mathit{psf}_\mathrm{\mathit{lpf}}(x,z)$), we can see that combining images with anisotropic resolution from multiple angles and superimposing them will create a reconstruction with isotropic spatial resolution. Since the center of the TF is over-emphasized, in practice we can apply a CT-like filtered backprojection algorithm to correct this bias. The theoretical PSF and TF of OCRT with full angular coverage are thus
\begin{equation}
\begin{split}
    \mathrm{\mathit{psf}}_\mathrm{\mathit{OCRT}}(x,z) &\propto
    \exp\left(-\frac{x^2+z^2}{\sigma_{z}^2}\right), \\
    H_\mathrm{\mathit{OCRT}}(k_x,k_z) &\propto
    \exp\left(-\frac{\sigma_z^2}{4}\left(k_x^2+k_z^2\right)\right),
\end{split}
\end{equation}
assuming that the original OCT axial resolution was superior to the lateral. We can also better appreciate more quantitatively why OCRT also obtains speckle reduction, as the speckle terms in the double summation in Eq. \ref{OCT_intensity} attenuate with angular diversity, as demonstrated in Eq. \ref{speckle_reduction} and Figs. \ref{fig:speckle_angle_compounded}. and \ref{fig:speckle_limited_angle}. 

\subsection{Comparison with coherent resolution enhancement in OCT}
While there are a few avenues for resolution enhancement in OCT, each has its own advantages and disadvantages, from practical and theoretical aspects. Coherent resolution enhancement techniques like ISAM are perhaps the most straightforward to implement in hardware, as they use the same setup and data acquisition pipeline as conventional OCT (unless higher NAs are desired). A major challenge, however, as discussed previously (Sec. \ref{coherent_approaches}), is maintaining phase stability among the lateral scans \cite{shemonski2014stability}, which is not required in conventional OCT. OCRT, however, does not require phase stability among its sequential measurements because OCRT uses intensity OCT images, in which the phase discarded. 

Another difference is that coherent methods, as synthetic aperture techniques, still maintain the band-pass nature of OCT and thus do not obtain speckle reduction (although the size of the speckle grain decreases in accordance to the expanded TF). OCRT, on the other hand, incoherently compounds the multi-angle images and takes advantage of the apparent low-pass nature of intensity OCT images to obtain speckle reduction. Thus, another perspective on this comparison between coherent and incoherent techniques is that both classes of techniques require additional information about the sample in 3D $k$-space, but utilize and synthesize this information differently. Synthetic aperture techniques maintain this information at their correct locations in $k$-space, while incoherent techniques like OCRT and speckle reduction rely on the discrete-sum assumption (Eqs. \ref{sum_assumption} and \ref{discrete_sum_k_space}), indicating that all pass-bands measure the same information about the sample with different speckle realizations (i.e., different observations of low-frequency beats caused by closely-spaced reflectors, as discussed Sec. \ref{k_space_speckle}). Therefore, these pass-bands can be demodulated to DC (i.e., by taking the amplitude squared) and superimposed to obtain speckle reduction and, if the pass-bands are anisotropic, resolution enhancement. A simple example of this distinction is that between using the full OCT bandwidth or multiple OCT bands to create a high-resolution image in the coherent case, and averaging multiple sub-bands to trade off axial resolution for speckle reduction in the incoherent case.

Finally, while synthetic aperture techniques place more emphasis on the angular spectral width, OCRT places more emphasis on the source spectral width and thus have different challenges. In particular, large source bandwidths are more susceptible to axial PSF broadening (equivalent to ultrafast pulse broadening) and thus require careful control of dispersion not only from the imaging system but also the sample (Sec. \ref{sample_induced_dispersion}). Similarly, large angular bandwidths are more susceptible to spatial aberrations and thus require carefully designed high-NA objectives and sometimes sample-induced aberrations \cite{adie2012computational,adie2012guide,Adie2015}. Finally, if both large source spectral bandwidths and angular bandwidths are desired, a situation more applicable to coherent resolution enhancement techniques, both aberrations and dispersion and their couplings would have to be addressed \cite{adie2012guide} (Sec. \ref{generalized_dispersion}).

\section{Conclusion and future directions}
In summary, we have advanced a full 3D $k$-space model of OCT, placing it in the context of general coherent imaging modalities. Using the Fourier diffraction theorem as the fundamental axiom on which the whole theory rests, we have derived the TFs of various implementations of OCT, including FF-OCT, LF-OCT, and point-scanning OCT, which are all band-pass TFs centered at $\mathbf{k}=(0,0,2k_0)$, assuming illumination in the $-k_z$ direction and collection in the $k_z$ direction. Conventional OCT processing ignores the curvature of the TFs, which originates from the Ewald sphere, that effectively couples the axial and lateral dimensions, thereby resulting in limited depths of focus. Using ISAM to resample the $k$-space coordinates in theory recovers the depth-invariant resolution promised by a 3D TF. Furthermore, as this $k$-space framework blurs the distinction between the axial and lateral dimensions, axial dispersion compensation and lateral aberration corrections may be unified as a generalized 3D pupil function, as is done in CAO. We also explained OCT speckle from the band-pass nature of the TF, and showed how angular compounding synthesizes incoherence. In doing so, we have shown that the intensity OCT image can be considered to be governed by a low-pass transfer function under the assumption that the sample is a discrete collection of scatterers. Based on this observation, we explained how OCRT simultaneously obtains speckle reduction and resolution enhancement. 

This unifying theoretical treatment of existing OCT techniques also highlights future research directions for the field. As discussed above, another relatively unexplored method of enhancing the lateral resolution of OCT is to coherently combine FF-OCT from multiple angles. Such an approach could have advantages over ISAM with high-NA objectives in terms of phase stability requirements and higher SNR away from the nominal focus. Such a strategy, along with FF-OCT with spatially incoherent illumination, can also be applied to transillumination OCT, for which all existing approaches have used point scanning. Another direction is alternative illumination and detection geometries that reduce the curvature of the OCT TF, thereby expanding the depth of focus without requiring resampling in $k$-space. Finally, while most of our discussions assumed single scattering in the first Born approximation, OCT would also benefit from deterministic modeling of multiple scattering in addition to statistical treatments, as scattering is a deterministic phenomenon for static samples. Doing so may extend the imaging range of OCT that is otherwise restricted by the first Born approximation, in the same way that using sophisticated scattering models recently advanced in the field of diffraction tomography has enabled transmission imaging of thicker, multiply scattering samples than previously possible \cite{kamilov2015learning,liu2017seagle, lim2017beyond, lim2019high, chowdhury2019high, pham2020three, chen2019multi}. While there have been efforts to create accurate wave-based forward models for OCT that model multiple scattering \cite{munro2015full, munro2016three}, they have yet to be used in inverse problem formulations to reconstruct the sample scattering potential or RI distribution. Accurate modeling of multiple scattering could not only extend the imaging depth of OCT, but also potentially reduce the cross-talk in FF-OCT with coherent illumination, which is caused by multiple scattering. Embracing optimization-based approaches to augment OCT is especially appropriate in this age where computational techniques are becoming much more feasible and commonplace.

In conclusion, we have presented a unified theoretical treatment of OCT that not only explains the fundamental concepts and properties of OCT, but also renders more transparent the connections among existing implementations of OCT as well as with other coherent imaging techniques. We  hope that this treatment will lead to new insights that encourage research in developing new OCT imaging techniques and extensions that yield ever more information about the sample under interrogation.

\section*{Acknowledgments}
National Institutes of Health (P30EY005722), National Science Foundation (CBET-1902904, DGF-1106401), Unrestricted Grant from Research to Prevent Blindness

\section*{Disclosures}
The authors have multiple patents related to various techniques discussed in this paper.

\bibliography{sample}

\end{document}